\documentclass[%
reprint,
superscriptaddress,
 amsmath,amssymb,
 aps,
]{revtex4-2}

\usepackage{graphicx}% Include figure files
\usepackage{dcolumn}% Align table columns on decimal point
\usepackage{bm}% bold math
\usepackage[caption=false]{subfig}
\usepackage{siunitx, soul, xcolor}
\usepackage{adjustbox}
\usepackage{gensymb}
\usepackage{bm}% bold math

\usepackage{xparse,xcoffins}

\ExplSyntaxOn
\NewCoffin\imagecoffin
\NewCoffin\labelcoffin

\keys_define:nn { miguel/label }
 {
  label   .tl_set:N = \l_miguel_label_tl,
  labelbox .bool_set:N = \l_miguel_label_box_bool,
  labelbox .default:n = true,
  fontsize .tl_set:N = \l_miguel_label_size_tl,
  fontsize .initial:n = \footnotesize,
  pos .choice:,
  pos/nw .code:n = \tl_set:Nn \l_miguel_label_pos_tl { left,up },
  pos/ne .code:n = \tl_set:Nn \l_miguel_label_pos_tl { right,up },
  pos/sw .code:n = \tl_set:Nn \l_miguel_label_pos_tl { left,down },
  pos/se .code:n = \tl_set:Nn \l_miguel_label_pos_tl { right,down },
  pos/n .code:n = \tl_set:Nn \l_miguel_label_pos_tl { hc,up },
  pos/w .code:n = \tl_set:Nn \l_miguel_label_pos_tl { left,vc },
  pos/s .code:n = \tl_set:Nn \l_miguel_label_pos_tl { hc,down },
  pos/e .code:n = \tl_set:Nn \l_miguel_label_pos_tl { right,vc },
  pos .initial:n = nw,
  unknown .code:n   = \clist_put_right:Nx \l_miguel_label_clist
                       { \l_keys_key_tl = \exp_not:n { #1 } }
 }
\clist_new:N \l_miguel_label_clist
\box_new:N \l_miguel_label_box
\box_new:N \l_miguel_label_image_box

\NewDocumentCommand{\xincludegraphics}{O{}m}
 {
  \group_begin:
  \tl_clear:N \l_miguel_label_tl
  \clist_clear:N \l_miguel_label_clist
  \keys_set:nn { miguel/label } { #1 }
  \tl_if_empty:NTF \l_miguel_label_tl
   {
    \miguel_includegraphics:Vn \l_miguel_label_clist { #2 }
   }
   {
    \SetHorizontalCoffin\imagecoffin
     {
      \miguel_includegraphics:Vn \l_miguel_label_clist { #2 }
     }
    \SetHorizontalCoffin\labelcoffin
     {
      \raisebox{\depth}
       {
        \bool_if:NTF \l_miguel_label_box_bool
         { \fcolorbox{white}{white}{\l_miguel_label_size_tl\l_miguel_label_tl} }
         { \l_miguel_label_size_tl\l_miguel_label_tl }
       }
     }
    \SetVerticalPole\imagecoffin{left}{-0pt+\CoffinWidth\labelcoffin/2}
    \SetVerticalPole\imagecoffin{right}{\Width-3pt-\CoffinWidth\labelcoffin/2}
    \SetHorizontalPole\imagecoffin{up}{\Height+10pt-\CoffinHeight\labelcoffin/2}
    \SetHorizontalPole\imagecoffin{down}{3pt+\CoffinHeight\labelcoffin/2}
    \use:x{\JoinCoffins\imagecoffin[\l_miguel_label_pos_tl]\labelcoffin[vc,hc]} 
    \TypesetCoffin\imagecoffin
   }
   \group_end:
 }
\NewDocumentCommand{\setlabel}{m}
 {
  \keys_set:nn { miguel/label } { #1 }
 }

\cs_new_protected:Nn \miguel_includegraphics:nn
 {
  \includegraphics[#1]{#2}
 }
\cs_generate_variant:Nn \miguel_includegraphics:nn { V }

\ExplSyntaxOff

\begin{document}
\preprint{APS/123-QED}

\title{Large spin-orbit torque in a-plane $\alpha$-Fe$_{2}$O$_{3}$/Pt bilayers}

\author{Igor Lyalin}
\affiliation{Department of Physics, The Ohio State University, Columbus, Ohio 43210, USA}
\author{Hantao Zhang}
\affiliation{Department of Electrical and Computer Engineering, University of California, Riverside, California 92521, USA}
\author{Justin Michel}
\affiliation{Department of Physics, The Ohio State University, Columbus, Ohio 43210, USA}
\author{Daniel Russell}
\affiliation{Department of Physics, The Ohio State University, Columbus, Ohio 43210, USA}
\author{Fengyuan Yang}
\affiliation{Department of Physics, The Ohio State University, Columbus, Ohio 43210, USA}
\author{Ran Cheng}
\affiliation{Department of Electrical and Computer Engineering, University of California, Riverside, California 92521, USA}
\affiliation{Department of Physics and Astronomy, University of California, Riverside, California 92521, USA}
\author{Roland K. Kawakami}
\email{kawakami.15@osu.edu}
\affiliation{Department of Physics, The Ohio State University, Columbus, Ohio 43210, USA}

\begin{abstract}
Realization of efficient spin-orbit torque switching of the N\'eel vector in insulating antiferromagnets is a challenge, often complicated by spurious effects.
Quantifying the spin-orbit torques in antiferromagnet/heavy metal heterostructures is an important first step towards this goal. 
Here, we employ magneto-optic techniques to study damping-like spin-orbit torque (DL-SOT) in a-plane $\alpha$-Fe$_2$O$_3$ (hematite) with a Pt spin-orbit overlayer.
We find that the DL-SOT efficiency is two orders of magnitude larger than reported in c- and r-plane hematite/Pt using harmonic Hall techniques.
The large magnitude of DL-SOT is supported by direct imaging of current-induced motion of antiferromagnetic domains that happens at moderate current densities. 
Our study introduces a new method for quantifying spin-orbit torque in antiferromagnets with a small canted moment and identifies a-plane $\alpha$-Fe$_2$O$_3$ as a promising candidate to realize efficient SOT switching.

\end{abstract}

\flushbottom
\maketitle
\thispagestyle{empty}

\begin{figure*}[th!]
\subfloat{\xincludegraphics[width=0.32\textwidth,label=(a)]{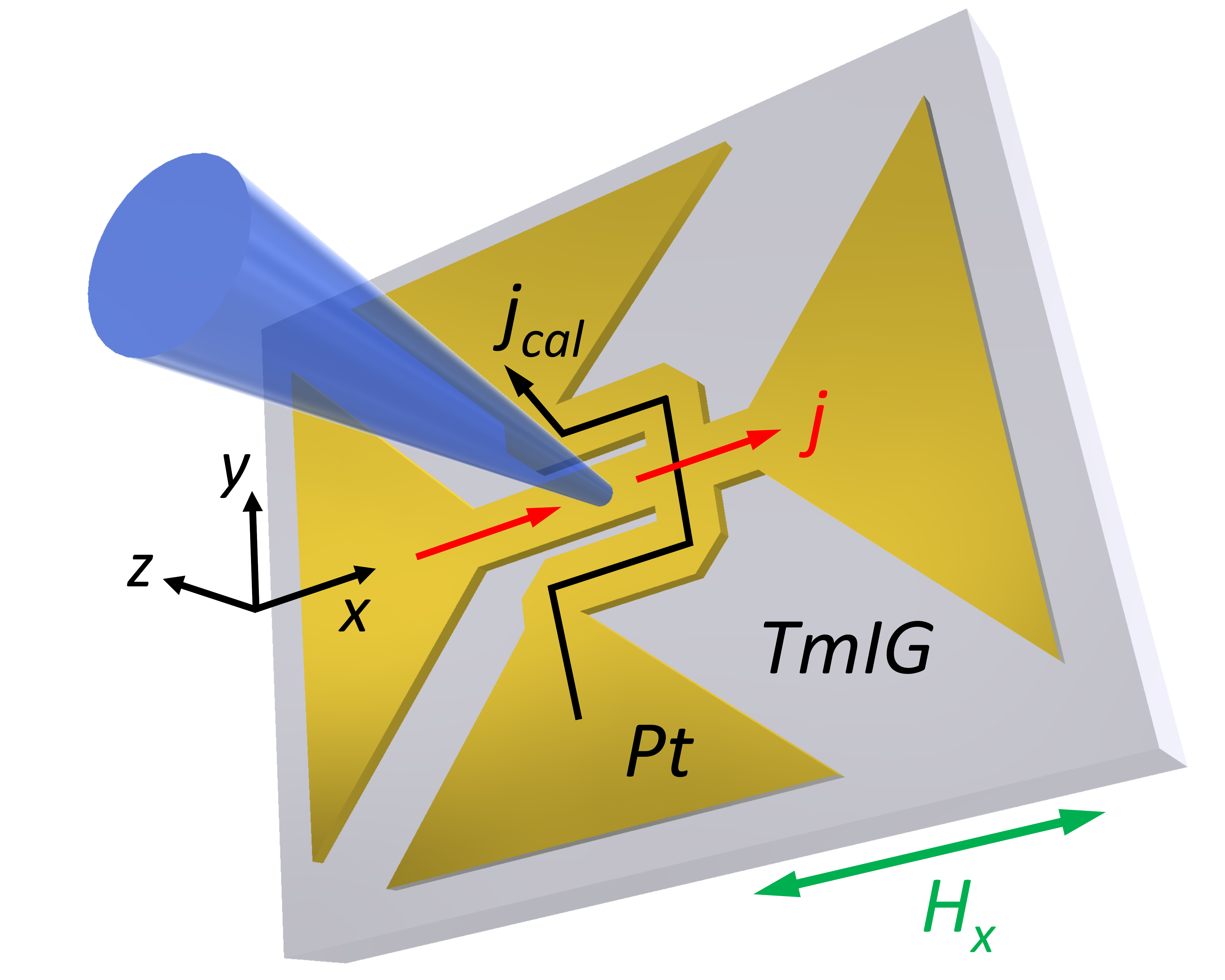}\label{fig:TmIG_experiment_schematics}}\hfill
\subfloat{\xincludegraphics[width=0.32\textwidth,label=(c)]{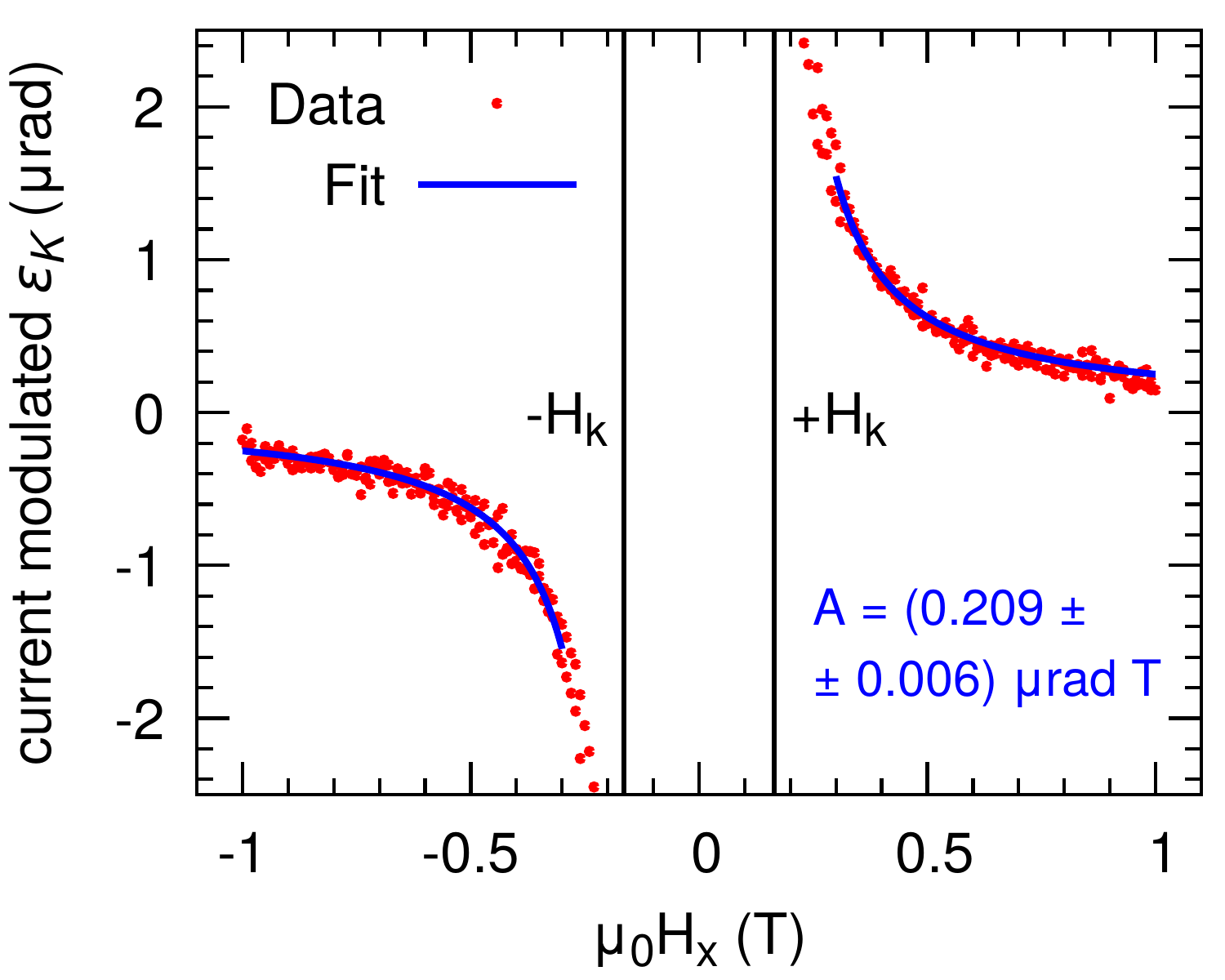}\label{fig:TmIG_KE_Hx_loop}}\hfill   
\subfloat{\xincludegraphics[width=0.32\textwidth,label=(e)]{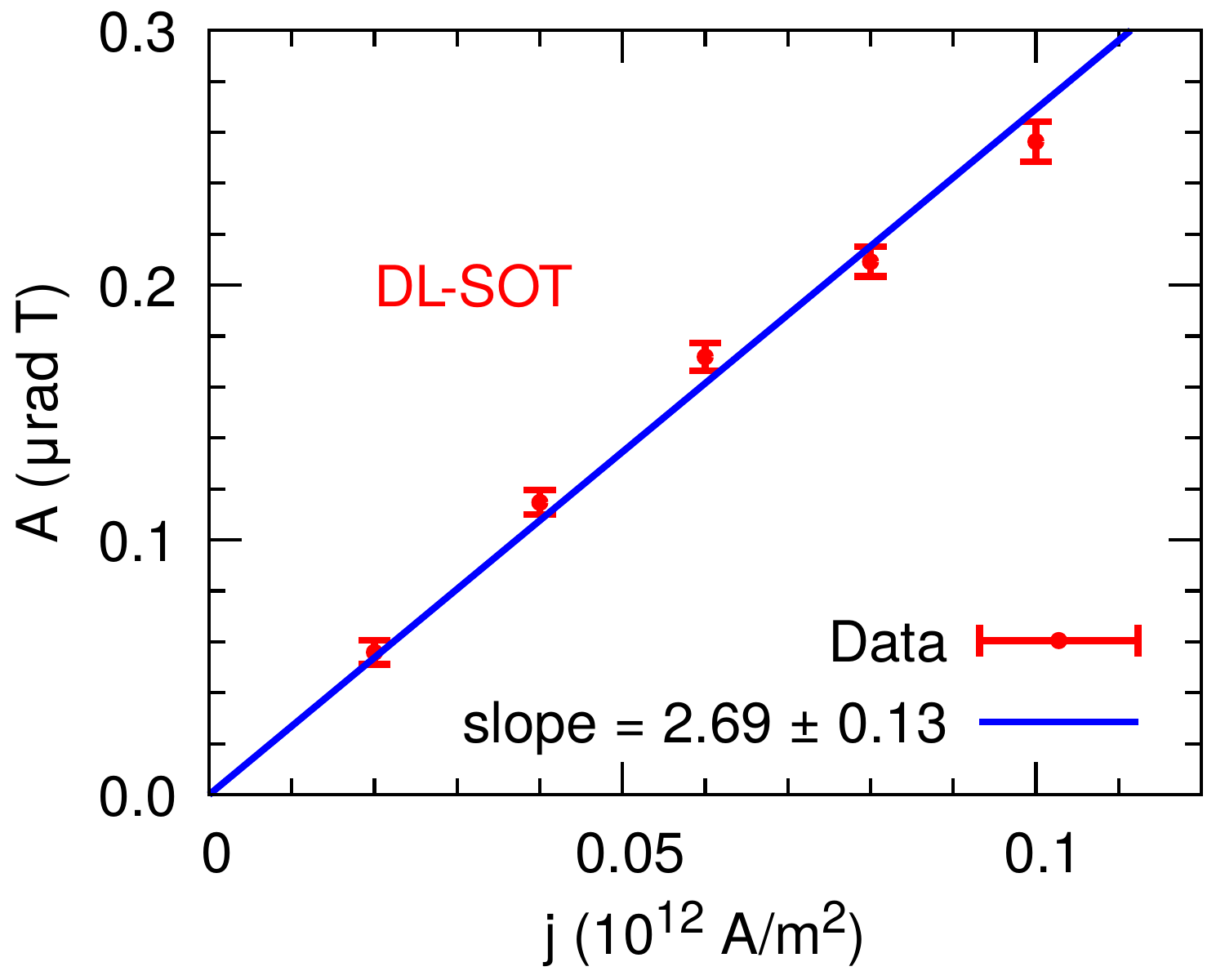}\label{fig:TmIG_current_modulated_KE_Hx_loop}}\hfill
\subfloat{\xincludegraphics[width=0.32\textwidth,label=(b)]{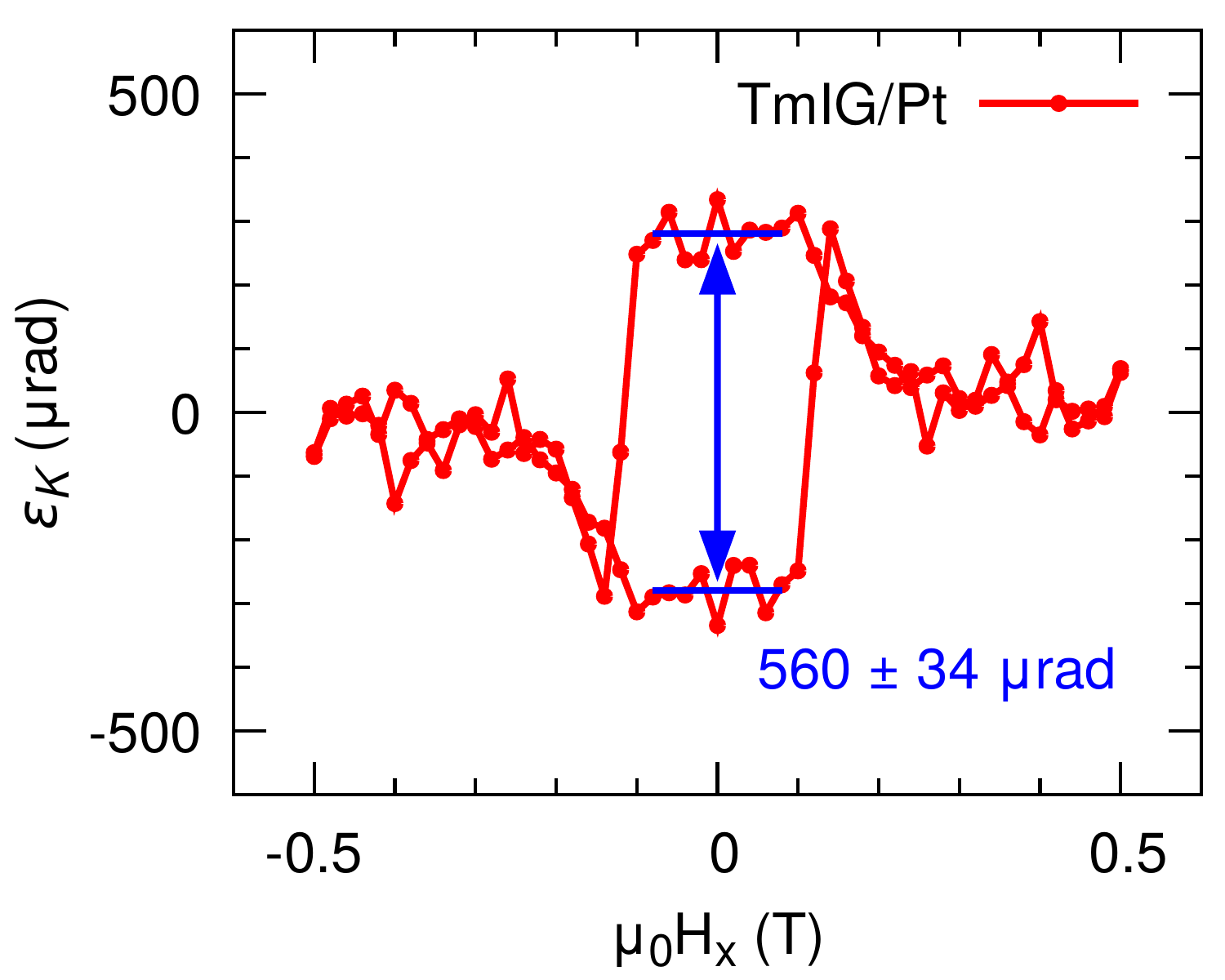}\label{fig:TmIG_current_modulated_KE_Hx_loop_calibration_wire}}\hfill 
\subfloat{\xincludegraphics[width=0.32\textwidth,label=(d)]{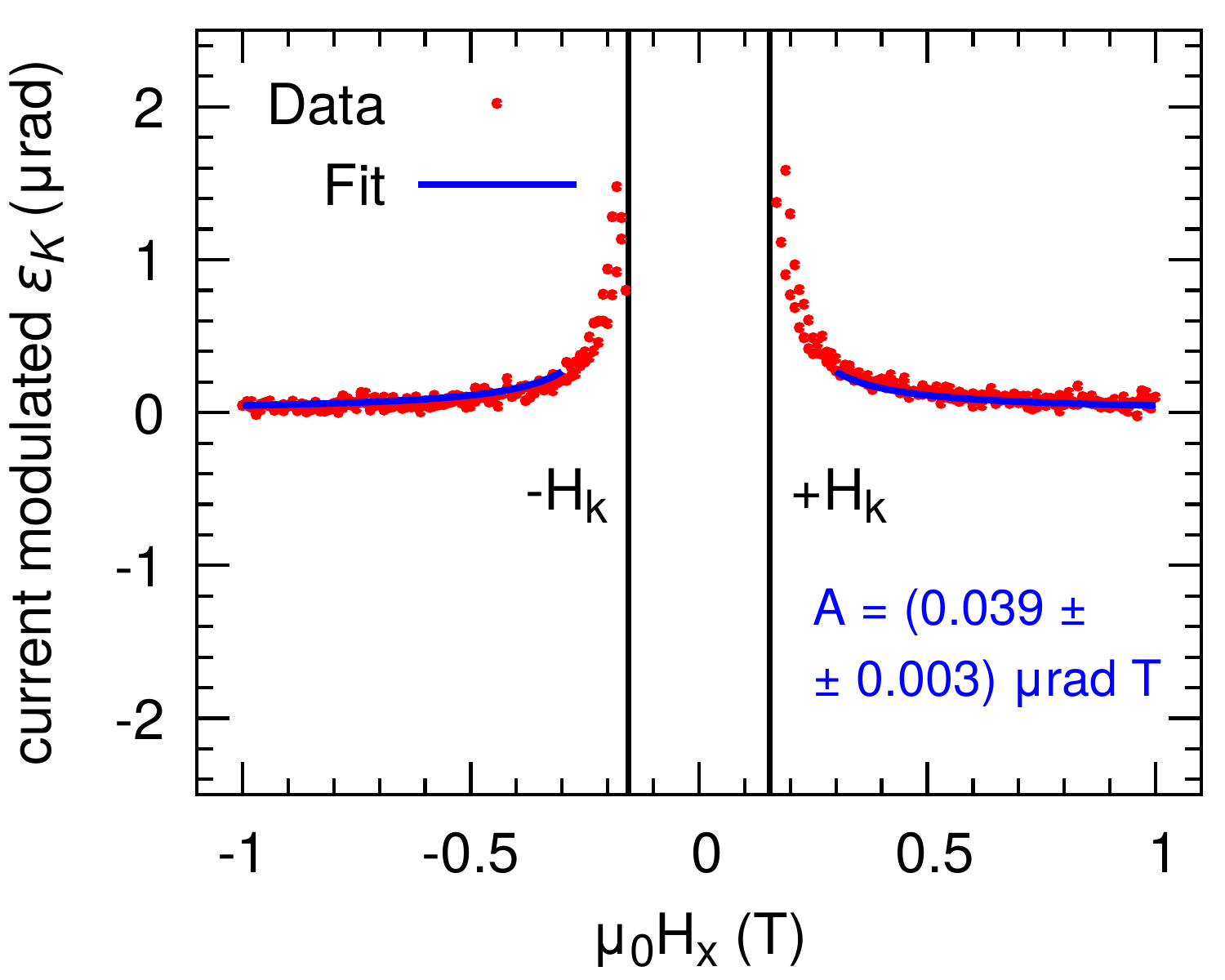} \label{fig:TmIG_current_modulated_KE_vs_j}}\hfill 
\subfloat{\xincludegraphics[width=0.32\textwidth,label=(f)]{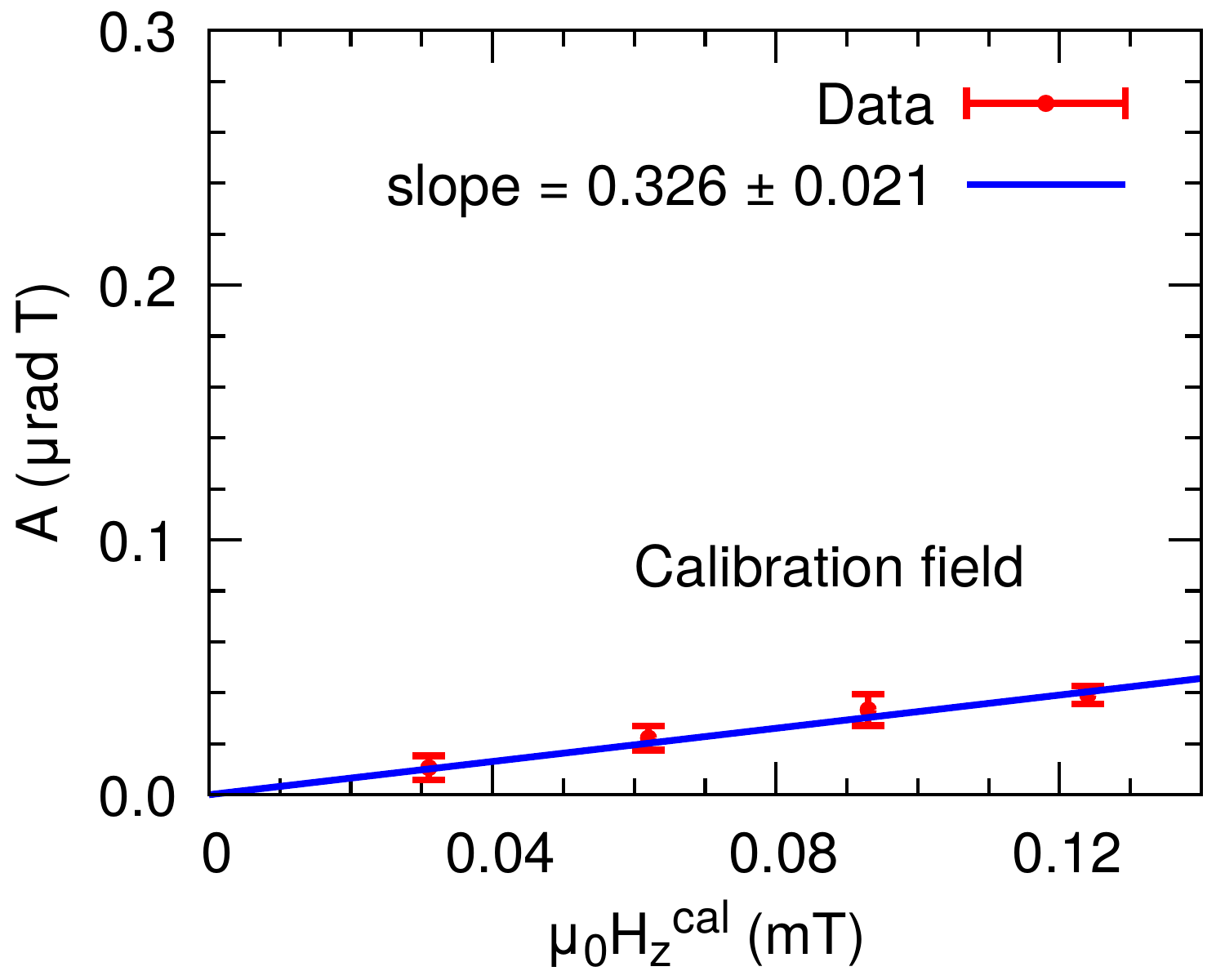}\label{fig:TmIG_calibration_KE_vs_OOP_Oe_Hz}}\hfill
\caption{TmIG/Pt (8/5\,nm):
  (a) Schematics of current modulated MOKE experiment.
  (b) Kerr ellipticity (KE) as a function of magnetic field $\mu_0H_x$ along hard-axis of TmIG.
  (c) Current-modulated KE as a function of magnetic field $\mu_0H_x$ for $j=8\times10^{10}$\,A/m$^2$.
  Here, the fitting parameter $A$ is proportional to the damping-like spin-orbit field $H_{DL}$.
  (d) Current-modulated KE as a function of magnetic field $\mu_0H_x$ for 8\,mA current applied through the calibration wire. 
  Here, the fitting parameter $A$ is proportional to the Oersted field.
  (e) Current-modulated KE as a function of current density $j$.
  (f) Current-modulated KE as a function of out-of-plane magnetic field $\mu_0H_z^{cal}$ due to the calibration wire.
  }
  \label{fig:TmIG_device}
\end{figure*}

%Recently, antiferromagnetic (AFM) materials have been attracting increasing attention due to advantages they can offer over conventional ferromagnetic (FM) materials.
%At the same time, the vanishing magnetization makes it challenging to detect and control AFMs.
Antiferromagnets (AFMs) display fast spin dynamics, produce no stray fields, and are robust against magnetic fields due to their vanishing magnetization.
Thus, in principle, they can offer faster and denser magnetic memory than spintronic devices based on conventional ferromagnetic (FM) materials~\cite{baltz_antiferromagnetic_2018,fukami_antiferromagnetic_2020,han_coherent_2023}.
Efficient electrical control of antiferromagnetic order is one of the holy grails of antiferromagnetic spintronics.
Spin-orbit torque (SOT) is an efficient way to manipulate magnetic order in ferromagnets~\cite{liu_spin-torque_2011,miron_perpendicular_2011,liu_spin-torque_2012,manchon_current-induced_2019}, however, effects of SOT on antiferromagnets are less studied.
In insulating AFM/heavy-metal (HM) bilayers, SOT switching is often complicated by spurious thermal effects arising from large electrical currents that are needed for the switching~\cite{chiang_absence_2019,baldrati_mechanism_2019,zhang_quantitative_2019,churikova_non-magnetic_2020,meer_direct_2021}.
Therefore, developing alternative methods for characterizing the SOT in AFM/HM is an important step towards understanding SOT physics in these materials and realizing efficient electrical control. 
Current-modulated magneto-optic Kerr effect (MOKE) is a powerful technique for characterizing SOT~\cite{fan_quantifying_2014,fan_all-optical_2016}.
Although this technique has been extensively applied to FMs~\cite{fan_quantifying_2014,fan_all-optical_2016,wang_anomalous_2019,lyalin_spinorbit_2021,pham_ferromagnetic_2023}, its application to SOT in AFMs has been mostly unexplored.

In this Letter, we report large damping-like SOT in antiferromagnetic a-plane $\alpha$-Fe$_{2}$O$_{3}$/Pt as characterized by current-modulated MOKE.  
The magneto-optic detection is facilitated by the presence of a small canted moment in $\alpha$-Fe$_{2}$O$_{3}$ generated by the Dzyaloshinskii–Moriya interaction. 
Indeed, we find that from the experimental geometry point of view, a-plane $\alpha$-Fe$_{2}$O$_{3}$ behaves similar to the case of a ferromagnetic material with perpendicular magnetic anisotropy (PMA).
To properly analyze current-modulated MOKE data, we develop a model that accounts for the antiferromagnetic nature of $\alpha$-Fe$_{2}$O$_{3}$.
%In previous works, c-plane [] and r-plane [] hematite/Pt heterostructes were studied using harmonic Hall technique [].
%Contrary to the case of FM/HM bilayers, the damping-like torque was found to be significantly smaller than the field-like torque.
Comparing the model with the experimental results, we find that the DL-SOT efficiency in a-plane hematite/Pt bilayers is two orders of magnitude larger than reported in c-plane and r-plane samples~\cite{cogulu_quantifying_2022,zhang_control_2022}.
Using MOKE microscopy, we directly observe current-induced motion of antiferromagnetic domains that happens at moderate current densities, which corroborates the large magnitude of the DL-SOT.
While our experimental geometry allows one to quantify only the damping-like component of the torque, we believe that this approach can be applied to other AFMs with a small canted moment.
Our study identifies a-plane $\alpha$-Fe$_2$O$_3$ as a promising candidate to realize efficient SOT switching and calls for the development of hematite thin films with a-plane orientation.

To quantify the SOT in a-plane $\alpha$-Fe$_{2}$O$_{3}$/Pt bilayers using current-modulated MOKE technique, we modify the calibration method that allows converting from MOKE units in radians to effective SOT field in Teslas~\cite{wang_anomalous_2019}, since calibration by a small Oersted field does not result in a measurable signal.
We first demonstrate the validity of our modified approach by applying it to a well-studied material, thulium iron garnet Tm$_3$Fe$_5$O$_{12}$ (TmIG), an insulating ferrimagnet  with perpendicular magnetic anisotropy.

TmIG/Pt films with 8/5\,nm thickness are grown by off-axis magnetron sputtering~\cite{ahmed_spin-hall_2019}
%on SGGG(111) substartes
and patterned into a device geometry shown in Fig.~\ref{fig:TmIG_experiment_schematics} by a combination of photolithography and argon ion milling.
The device channel width is 20\,$\mu$m.
We detect the out-of-plane magnetization, $m_z$, using the polar MOKE geometry and measuring the Kerr ellipticity $\varepsilon_K \sim m_z$ of a linearly polarized laser light reflected from the sample. 
The laser beam is focused on the sample surface to $\sim$\,5\,$\mu$m spot size, the wavelength of the light is 400\,nm.
Fig.~\ref{fig:TmIG_KE_Hx_loop}
%\red{[the labels were wrong. They go row by row regardless of the label designation. This was 1d and should be 1b. There were many other errors. You need to arrange the labels in the ``wrong" way to get the correct referencing. It's confusing so made the changes for you.]} 
shows a Kerr ellipticity (KE) hysteresis loop with external magnetic field $\mu_0H_x$ applied along hard-axis of TmIG.
When $\mu_0H_x$ is larger than the anisotropy field $\mu_0H_k$, the TmIG magnetization is along $x$ direction.
With the current applied along the $x$ axis as well, the spin Hall effect (SHE) in the Pt layer generates spin current with polarization $\bm \sigma$ along $y$ axis.
In this geometry, the effective damping-like field ${\bm H_{DL}} \sim [\bm{m} \times \bm{\sigma}] \sim [\bm{\hat{x}} \times \bm{\hat{y}}]$ results into a small tilt of magnetization towards the sample normal, thus inducing an out-of-plane component $\Delta m_z$.
For $|H_x| > |H_k|$ and a small effective damping-like field $H_{DL} \ll |H_x - H_k|$ the out-of-plane current-induced magnetization can be written as:
\begin{equation}
\dfrac{\Delta m_z}{m} = \dfrac{\Delta \varepsilon_K^j}{\varepsilon_K^{max}} = \dfrac{\mu_0 H_{DL}}{|\mu_0 H_x - \mu_0 H_k|}
 \label{eq:KE_normalization}
\end{equation}
where $\varepsilon_K^{max}$ is Kerr ellipticity that corresponds to the magnetization $m$ being fully out-of-plane.

The current-modulated Kerr ellipticity $\Delta \varepsilon_K^j$ is detected using a combination of quarter-wave plate, half-wave plate, Wollaston prism, balanced photodetector and lock-in amplifier at the frequency of the applied ac current.
Fig.~\ref{fig:TmIG_current_modulated_KE_Hx_loop} shows  $\Delta \varepsilon_K^j$ as a function of magnetic field $\mu_0H_x$ for a current density of $j=8\times10^{10}$\,A/m$^2$.
Following eq.~(\ref{eq:KE_normalization}), the data can be fitted using expression $\Delta \varepsilon_K^j = A/ | \mu_0 H_x - \mu_0 H_k|$ with two fitting parameters $A$ and $H_k$, where $A=\varepsilon_K^{max} \mu_0 H_{DL}$.
Measuring current-modulated KE at different current densities, we verify that $A \sim H_{DL}$ scales linearly with $j$, as expected for the effective SOT field.
The current density dependence is plotted in Fig.~\ref{fig:TmIG_current_modulated_KE_vs_j}.
Using the $A/j$ slope normalized by $\varepsilon_K^{max}$, extracted as a half of the hysteresis loop opening at $\mu_0H_x = 0$ in Fig.~\ref{fig:TmIG_KE_Hx_loop}, one can estimate the DL-SOT efficiency:
\begin{equation}
    \dfrac{A}{j} \dfrac{1}{\varepsilon_K^{max}} = \dfrac{\mu_0 H_{DL}}{j} = \dfrac{\hbar}{2 e} \dfrac{\xi_{DL}}{ M_s t_{FM}}
    \label{eq:H_eff_DL-SOT}
\end{equation}
Using the measured values $A/j = 2.69 \pm 0.13 \,\mu$rad\,T per 10$^{12}$ A/m$^2$, $\varepsilon_K^{max} = 280 \pm 17\, \mu$rad, and  $t_{FM} = 8$\,nm along with $M_s = 110$\, kA/m from the literature (saturation magnetization of TmIG~\cite{ahmed_spin-hall_2019}), we estimate the effective DL-SOT field and efficiency to be $\mu_0 H_{DL}/j = 9.6 \pm 0.8$\,mT per 10$^{12}$ A/m$^2$ and $\xi_{DL} = 0.026 \pm 0.002$, respectively.
This agrees well with studies on insulating FM/HM bilayers, in which DL-SOT efficiency have been found to be smaller than in metallic FM/HM.
Literature on TmIG/Pt reports values of $0.01-0.03$ SOT efficiency~\cite{avci_current-induced_2017,li_giant_2023}.
We note that to correctly estimate $\varepsilon_K^{max}$, the magnetic material should be in a single domain state within the laser spot size (at $\mu_0 H_x = 0$) or the out-of-plane $\mu_0 H_z$ hysteresis loop should be measured to extract $\varepsilon_K^{max}$.
\begin{figure*}[ht!]
\hspace{0cm}\subfloat{\xincludegraphics[width=0.32\textwidth,label=(a)]{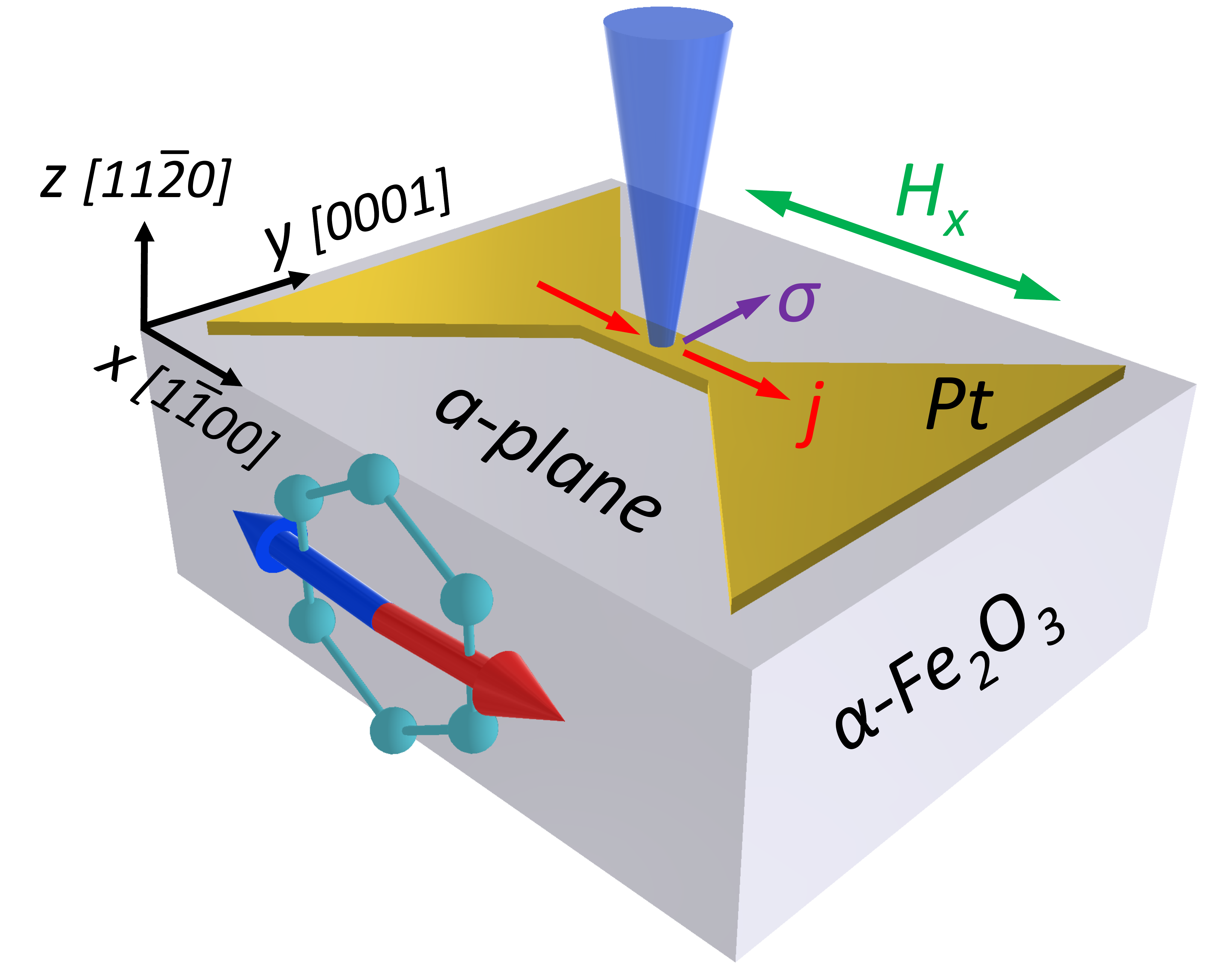}\label{fig:Fe2O3_experiment_schematics}}\hspace{0.5cm}
\subfloat{\xincludegraphics[width=0.49\textwidth,label=(b)]{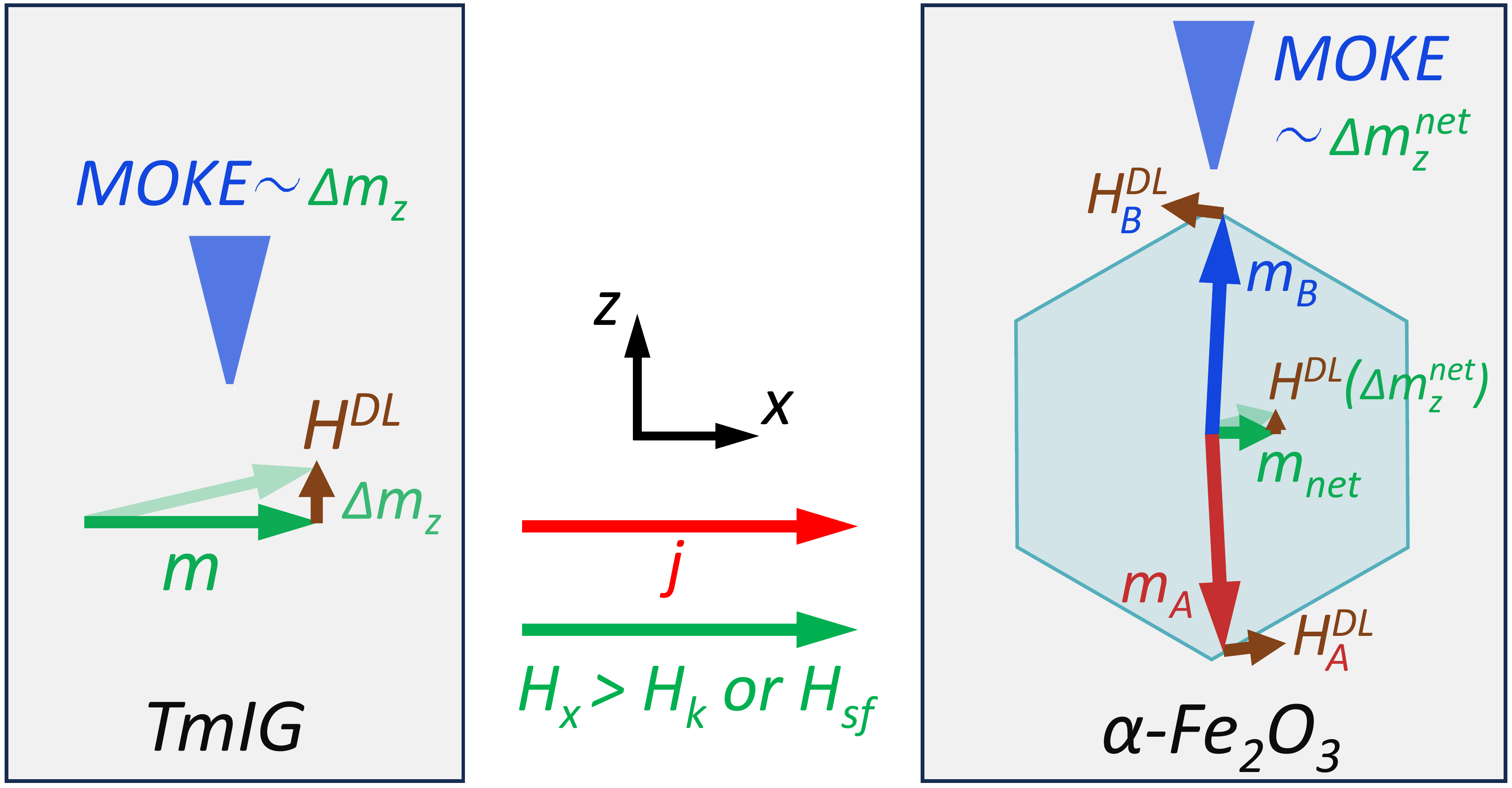}\label{fig:SOT_schematics_TmIG_and_Fe2O3}}\hspace{1cm}
\subfloat{\xincludegraphics[width=0.33\textwidth,label=(c)]{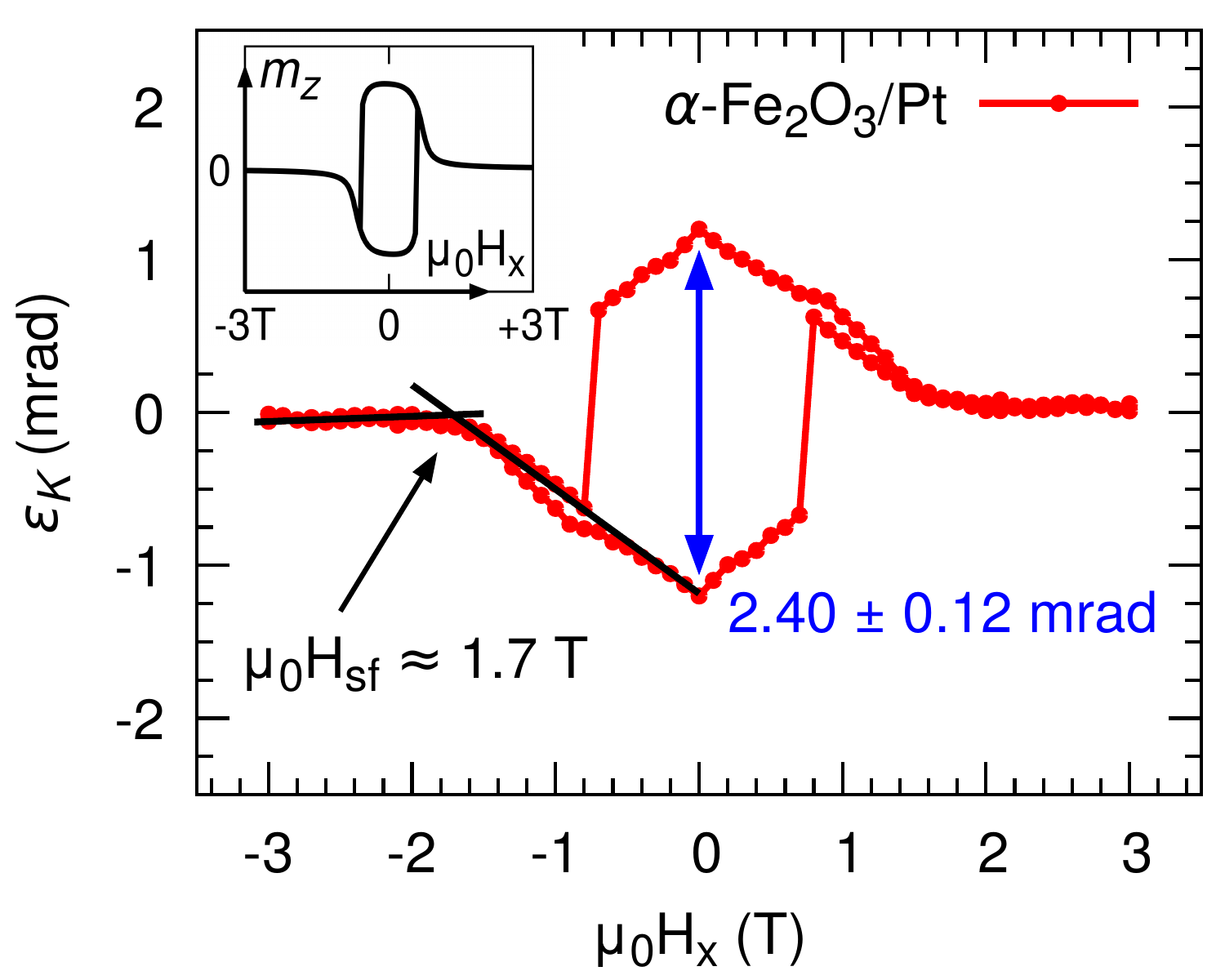}\label{fig:Fe2O3_KE_Hx_hloop_1st_device}}\hfill
\subfloat{\xincludegraphics[width=0.33\textwidth,label=(d)]{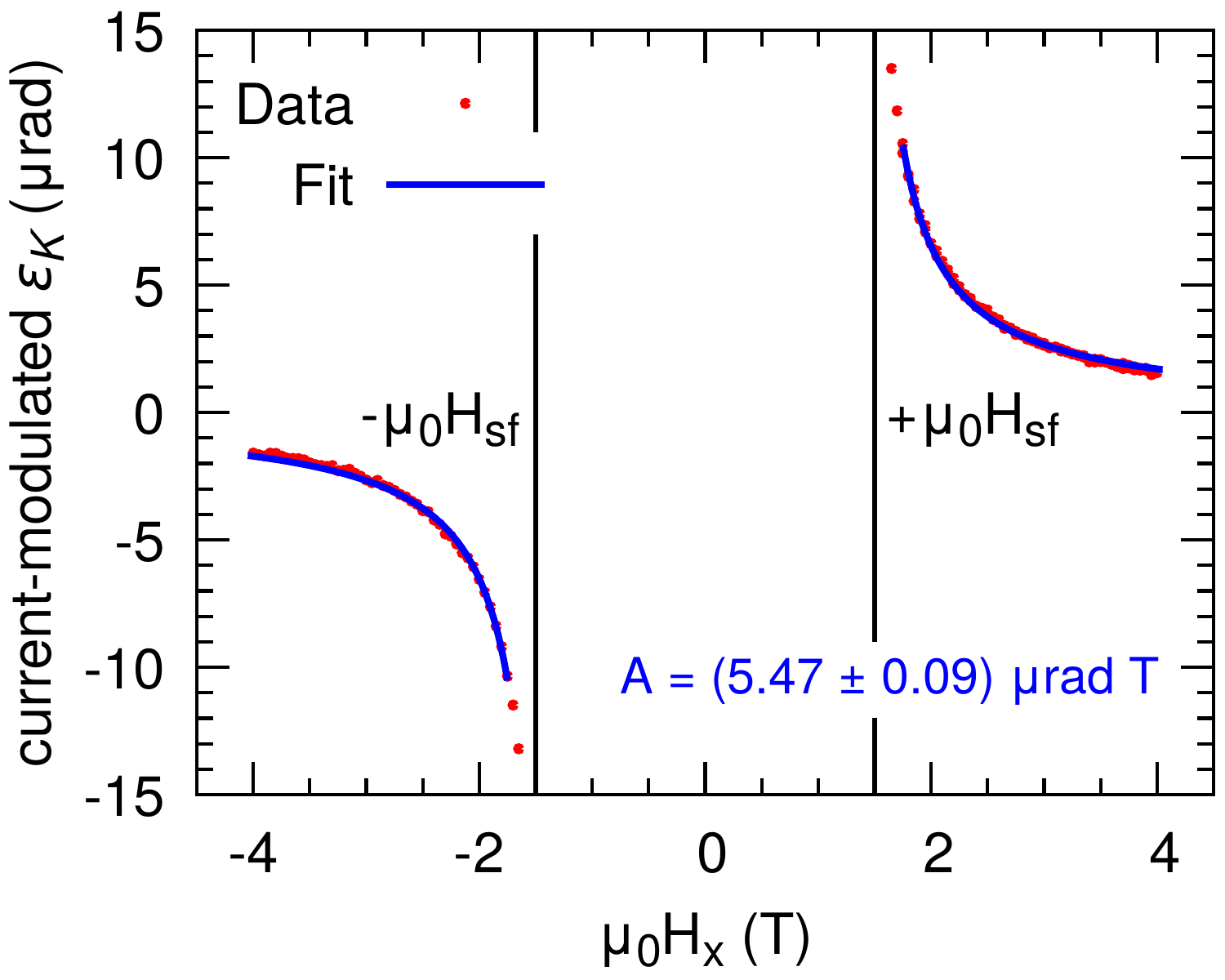}\label{fig:Fe2O3_SOT_KE_Hx_hloop_1st_device}}\hfill
\subfloat{\xincludegraphics[width=0.33\textwidth,label=(e)]{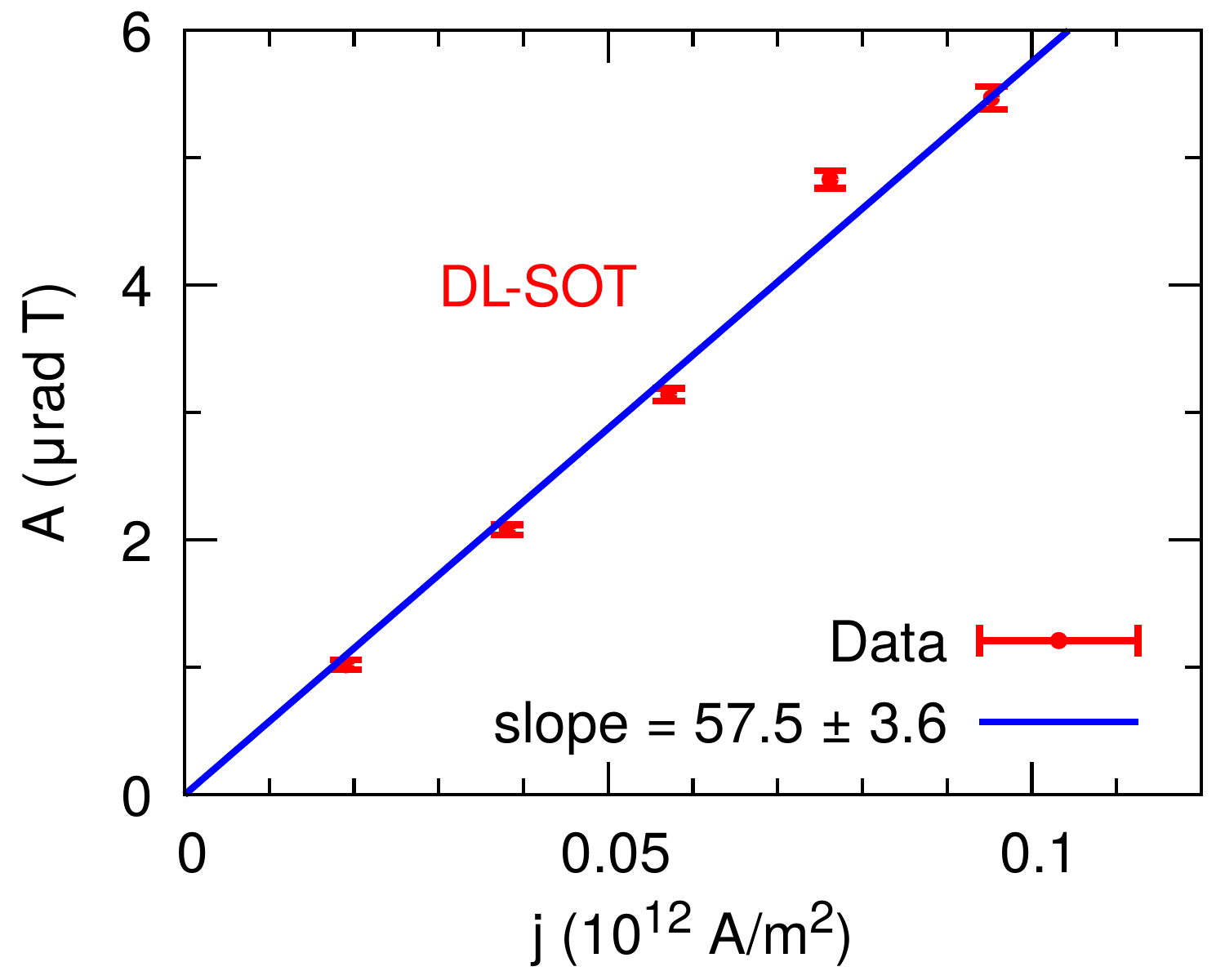}\label{fig:Fe2O3_KE_vs_j_1st_device}}\hfill
\caption{a-plane $\alpha$-Fe$_{2}$O$_{3}$/Pt (bulk/5\,nm):
  (a) Schematics of the current modulated MOKE experiment.
  (b) Schematics of DL-SOT in TmIG vs  $\alpha$-Fe$_{2}$O$_{3}$.
  (c) Kerr ellipticity as a function of magnetic field $\mu_0H_x$ along the [1$\bar1$00] direction. The inset shows a simulated $m_z$ vs $\mu_0H_x$ loop.
  (d) Current-modulated KE as a function of magnetic field $\mu_0H_x$ for $j=9.5\times10^{10}$\,A/m$^2$. 
  The red dots are the experimental data, and the blue curve is a fit to the theoretical model, which yields a best-fit value of $A = 5.47\,\mu$rad T which is proportional to the damping-like spin-orbit field $H_{DL}$ (see text for the definition of $A$).
  (e) Fitting parameter $A$ $(\sim H_{DL})$ as a function of current density $j$.
  }
  \label{fig:Fe2O3_1st_device}
\end{figure*}

\begin{figure*}[ht!]
\subfloat{\xincludegraphics[width=0.9\textwidth,label=]{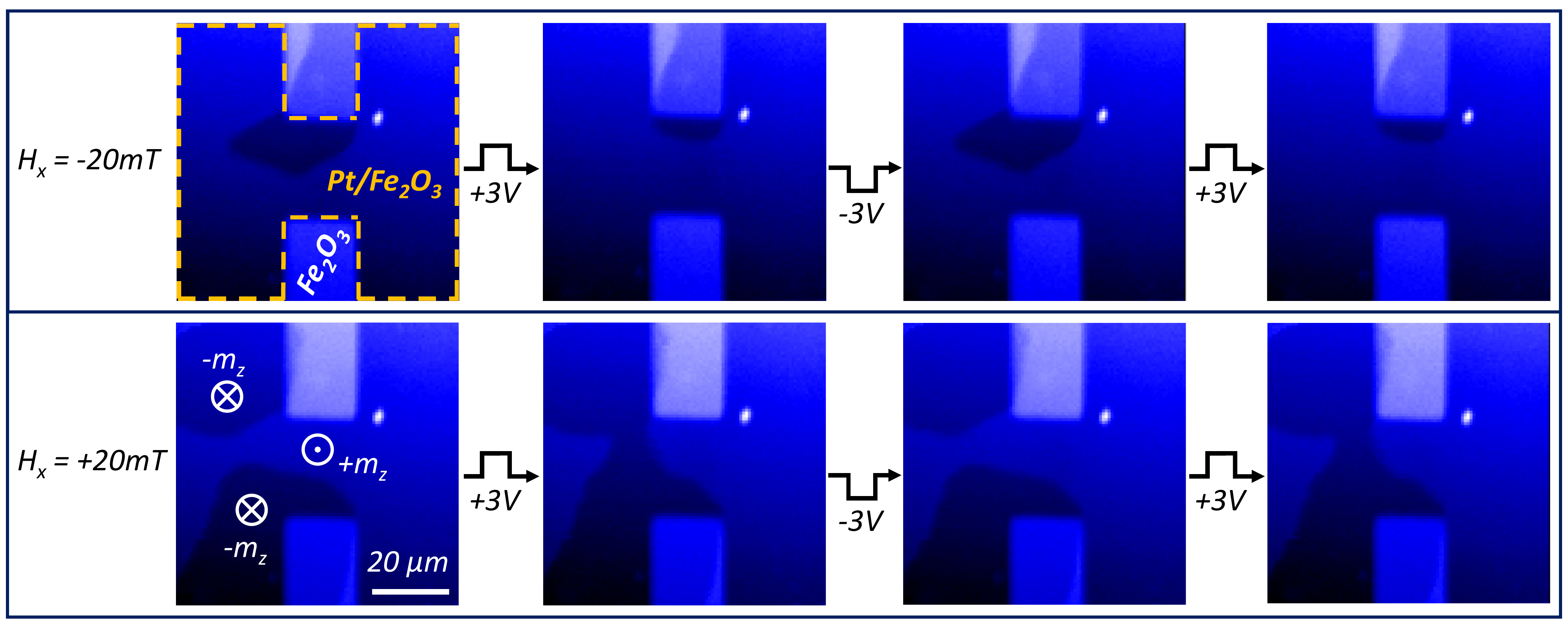}\label{fig:Fe2O3_MOKE_images_pulsing}}\hfill
\caption{MOKE microscopy maps of an $\alpha$-Fe$_{2}$O$_{3}$/Pt device revealing current-induced domain motion. 
The 3\,V amplitude of voltage pulses corresponds to $j = 2.4 \times 10^{11}$\,A/m$^2$ current density. 
The pulse duration is 10\,$\mu$s. 
A white spot in the top right quadrant is due to a dust particle on the Pt surface.}
  \label{fig:domain_switching}
\end{figure*}

While the TmIG/Pt results agree well with reported values, we verify the validity of the experimental approach further using a calibration of effective DL-SOT field by Oersted field generated by a wire of a known geometry~\cite{fan_quantifying_2014,wang_anomalous_2019}.
The current path for the calibration wire in our device is shown in Fig.~\ref{fig:TmIG_experiment_schematics} and is labeled as $\bm{j}_{cal}$.
Fig.~\ref{fig:TmIG_current_modulated_KE_Hx_loop_calibration_wire} shows the current-modulated KE response to an out-of-plane Oersted field generated by a calibration wire.
Unlike the effective DL-SOT field, ${\bm H_{DL}} \sim [\bm{m} \times \bm{\sigma}]$, that flips sign upon magnetic field $\mu_0 H_{x}$ reversal (Fig.~\ref{fig:TmIG_current_modulated_KE_Hx_loop}), the Oersted field is independent of $\bm m$.
Thus, the Oersted-field-induced MOKE signal has even symmetry with respect to the magnetic field, as demonstrated in Fig.~\ref{fig:TmIG_current_modulated_KE_Hx_loop_calibration_wire}.
Using Ampere's law, we calculate the magnetic field generated by the wire (drawn in Fig.~\ref{fig:TmIG_experiment_schematics}).
For the maximal applied current of 8\,mA, the COMSOL simulation yields a magnetic field of 0.124\,mT.
Thus we can calibrate the MOKE signal due to the effective DL-SOT against the Oersted field. 
Fig.~\ref{fig:TmIG_calibration_KE_vs_OOP_Oe_Hz} shows the current-modulated MOKE signal as a function of the Oersted field generated by current in the calibration wire.
Using the slopes in Fig.~\ref{fig:TmIG_current_modulated_KE_vs_j} and~\ref{fig:TmIG_calibration_KE_vs_OOP_Oe_Hz}, we find $\mu_0 H_{DL}/j = 8.3 \pm 0.7$\,mT per 10$^{12}$~A/m$^2$, which within experimental errors agrees with the value estimated by the $\varepsilon_K^{max}$ normalization method.
We note that the $\varepsilon_K^{max}$ normalization method might be advantageous in the situation where the Oersted field is not strong enough to induce a reliable  MOKE response, which we have found to be the case for our $\alpha$-Fe$_{2}$O$_{3}$/Pt devices.

Having verified that our modified approach is valid, we discuss the spin-orbit torque measurements in a-plane $\alpha$-Fe$_{2}$O$_{3}$/Pt (bulk/5\,nm).
A commercial bulk crystal with dimensions $5\times5\times1$\,mm and  (11$\bar{2}$0) cut (a-plane) is purchased from MTI Corporation.
5\,nm Pt is deposited by magnetron sputtering at room temperature.
%Next, we perform MOKE measurements on $\alpha$-Fe$_2$O$_3$/Pt to quantify DL-SOT.
The geometry of the current-modulated MOKE experiment, Fig.~\ref{fig:Fe2O3_experiment_schematics}, is similar to the measurements on TmIG/Pt, with a difference that now the spin-flop field, $H_{sf}$, is needed to align the magnetization along $x$ direction, instead of the anisotropy field $H_k$.
Fig.~\ref{fig:SOT_schematics_TmIG_and_Fe2O3} illustrates the difference between the action of damping-like spin-orbit torque on ferrimagnetic TmIG and antiferromagnetic $\alpha$-Fe$_2$O$_3$. 
Charge current through the Pt layer is applied along the [1$\bar{1}$00] direction ($x$-axis in Fig.~\ref{fig:Fe2O3_experiment_schematics}) and as a result, the spin polarization $\bm \sigma$ is along [0001] ($y$-axis).
The effective damping-like field acts on the magnetic moments of both sublattices: $\bm H^A_{DL} \sim [\bm m_A \times \bm \sigma]$ and $\bm H^B_{DL} \sim [\bm m_B \times \bm \sigma]$.
The two torques act constructively to rotate the N\'eel vector and the canted magnetic moment ($\bm m_{net}$) in the basal plane.
For $|H_x| > |H_{sf}|$, the $\bm m_{net}$ is rotated out of the sample plane and the resultant out-of-plane component $\Delta m^{net}_z$ can be detected by current-modulated polar MOKE.

Kerr ellipticity as a function of magnetic field swept along [1$\bar{1}$00] direction ($H_x$) is plotted in Fig.~\ref{fig:Fe2O3_KE_Hx_hloop_1st_device}.
Somewhat surprisingly, the KE signal due to the small canted moment in antiferromagnetic $\alpha$-Fe$_2$O$_3$ is large, $\varepsilon_K^{max} = 1.20 \pm 0.06$\,mrad, roughly $4-5$ times larger than for ferrimagnetic TmIG. 
This could be explained by the larger thickness of $\alpha$-Fe$_2$O$_3$ vs TmIG (bulk vs 8\,nm) and the difference in magneto-optic coefficients at 400\,nm.
The spin-flop field can be estimated from KE hysteresis loop along [1$\bar{1}$00], as shown in Fig.~\ref{fig:Fe2O3_KE_Hx_hloop_1st_device}, $\mu_0 H_{sf} \approx 1.7$\,T, which agrees with an estimation based on spin Hall magnetoresistance measurements, $\mu_0 H_{sf} \approx 1.5$\,T, see SM Sec. S2~\cite{SupplMat} for details.

We also simulate $m_z$ vs $\mu_0 H_x$ numerically (see SM Sec. S4~\cite{SupplMat} for details).
Setting the exchange field $\mu_0 H_{E} = 900$\,T, anisotropy field in the basal plane $\mu_0 H_K = 6 \times 10^{-4}$\,T, DMI field $\mu_0 H_D = 2.5$\,T, which are close to values reported in the literature~\cite{williamson_antiferromagnetic_1964,mizushima_effective_1966,elliston_some_1968,lebrun_anisotropies_2019,wang_spin_2021}, and a misalignment angle of $3\degree$ between the external magnetic field and the [1$\bar{1}$00] crystallographic direction, we are able to qualitatively reproduce the experimental data.
The simulated hysteresis loop is shown in the inset of Fig.~\ref{fig:Fe2O3_KE_Hx_hloop_1st_device}.
We note that an exact shape of the hysteresis loop is sensitive to the misalignment angle of the magnetic field.
%between the external magnetic fieldof $3\degree$ and the [1$\bar{1}$00] crystallographic direction.
%which could explain the slight difference in $H_{sf}$ values.

The current-modulated Kerr ellipticity as a function of magnetic field $\mu_0H_x$ for a current density of $j=9.5\times10^{10}$\,A/m$^2$ is shown in Fig.~\ref{fig:Fe2O3_SOT_KE_Hx_hloop_1st_device}.
We show that for $|H_x| > |H_{sf}|$, the data can be fitted using an expression (see SM Sec. S4~\cite{SupplMat}):
%$\Delta \varepsilon_K^j = A / | \mu_0 H_x - \mu_0 H_{sf}|$.
\begin{equation}
\dfrac{\Delta m_z}{m} = \dfrac{\Delta \varepsilon_K^j}{\varepsilon_K^{max}} = \dfrac{\mu_0 H_{DL}}{\left|\mu_0 H_x - \mu_0 \dfrac{4 H_{E} H_K}{( H_x + H_D)}\right|}
 \label{eq:KE_normalization}
\end{equation}
%where $H_E$ is the exchange field, $H_K$ is the anisotropy field, $H_D$ is the DMI field.
We fix $\mu_0 H_{D}$ to a value of 2.5\,T as it is more established in the literature~\cite{williamson_antiferromagnetic_1964,mizushima_effective_1966, elliston_some_1968,lebrun_anisotropies_2019,wang_spin_2021} and fit the current-modulated Kerr ellipticity data $\Delta \varepsilon_K^j (H_x)$ using two fitting parameters $A=\varepsilon_K^{max} \mu_0 H_{DL}$ and $B = \mu_0^2 H_{E} H_K$.
The result of the fit is the blue curve in Fig.~\ref{fig:Fe2O3_SOT_KE_Hx_hloop_1st_device}.

Next, we measure the current-modulated MOKE signal at different current densities and repeat the fitting procedure. 
As shown in Fig.~\ref{fig:Fe2O3_KE_vs_j_1st_device}, we observe that $A$ scales linearly with current density $j$, consistent with SOT origin of the MOKE signal.
Similar to TmIG/Pt analysis, using the measured values of the slope $A/j = 57.5 \pm 3.6 \,\mu$rad\,T per 10$^{12}$ A/m$^2$ and $\varepsilon_K^{max} = 1.20 \pm 0.06$\,mrad, we can estimate the effective DL-SOT field.
We find $\mu_0 H_{DL}/j = 47.9 \pm 3.9$\,mT per 10$^{12}$\,A/m$^2$.
We note that the fitting is not too sensitive to the fixed value of the DMI field.
Varying the $\mu_0 H_D$ value in the $1-10$\,T range changes the value of $\mu_0 H_{DL}$ only by 20\%. 
%\red{The fitted value of $B$ yields $\mu_0^2 H_E H_K = [1.16-1.31]$\,T$^2$, which is within xxx\% of the value used for plotting the inset of Figure [2c] ($\mu_0^2 H_E H_K = 0.54$\,T$^2$), which demonstrates that the fitting yields reasonable values.} 

%\blue{[need new equation to translate $H_{DL}$ to DL-SOT efficiency $\xi_{DL}$? If not, how to explain 200\% efficiency?]}

The $\mu_0 H_{DL}/j$ value is two orders of magnitude larger than effective DL-SOT field reported in c-plane and r-plane hematite using the harmonic Hall technique~\cite{cogulu_quantifying_2022,zhang_control_2022} and larger than values for thick metallic FM/Pt bilayers~\cite{liu_spin-torque_2011}.
We note that we study a bulk sample and that even larger $H_{DL}$ should be expected for thin films.
We also point out that in c-plane $\alpha$-Fe$_2$O$_3$, $\bm H_{DL}$ points along the c-axis and thus it needs to overcome the strong easy-plane anisotropy in order to realize the switching.
For r-plane orientation, this effect is mitigated, while for a-plane orientation, $\bm H_{DL}$ fully lies within the easy-plane, and consequently, its effect on the rotation of the N\'eel vector is maximized.
%Although, the effect of the anisotropy is included in the calculations, it is possible that the suppressed raw signal in combination with spurious effects could lead to incorrect estimation of the magnitude of the DL-SOT.  
%t_{FM} = 24.6$\,nm (light penetration depth of $\alpha$-Fe$_{2}$O$_{3}$/Pt at 400\,nm from absorption coefficient in literature [])
Thus, a-plane $\alpha$-Fe$_2$O$_3$/Pt might be a promising material system to realize efficient SOT switching of the N\'eel vector.
To check that our results are robust to a different location on the sample, device geometry, and optical alignment, we reproduce results on a different $\alpha$-Fe$_2$O$_3$/Pt device. 
The data on the other device are shown in the SM Sec.~S3~\cite{SupplMat}.
We find $\mu_0 H_{DL}/j = 54.1 \pm 3.2$\,mT per 10$^{12}$\,A/m$^2$, which agrees well with the result on the device shown in Fig.~\ref{fig:Fe2O3_1st_device}.

%One can notice that the shape of the hard axis hysteresis loop is different between device 1 and 2 which we assign to magnetic domains of Fe$_{2}$O$_{3}$ being larger or comparable to the laser beam spot size in our experiment $\sim 5\,\mu$m.

Finally, we utilize MOKE microscopy to directly image AFM domains in a-plane $\alpha$-Fe$_2$O$_3$/Pt.
Fig.~\ref{fig:domain_switching} shows polar MOKE maps of a $21\,\mu$m wide and $20\,\mu$m long device.
Large, tens of micrometers in size, AFM domains with a canted moment $\bm m_{net}$ along $+z$ (light blue) and $-z$ (dark blue) direction are observed.
Using a small bias field, $\mu_0 H_x = \pm 20$\,mT, we demonstrate current-driven AFM domain motion.
For $\mu_0 H_x = -20$\,mT, a $-m_z$ domain in the center of the device reproducibly shrinks (expands) after a positive $+3$\,V (negative $-3$\,V) pulse that corresponds to a moderate current density of $2.4 \times 10^{11}$\,A/m$^2$.
The duration of pulses is 10\,$\mu$s.
When the bias field is reversed to $\mu_0 H_x = -20$\,mT, the polarity of the domain motion changes.
Now, a $+m_z$ domain shrinks (expands) after a positive $+3$\,V (negative $-3$\,V) pulse.
This behavior is fully consistent with a protocol for damping-like SOT-driven domain motion in FM/HM bilayers~\cite{velez_high-speed_2019}.
The SOT-driven motion of AFM domain at moderate current densities supports the large measured value of DL-SOT efficiency.
We note that for many AFMs, synchrotron facilities are needed for direct imaging of AFM domains~\cite{baldrati_mechanism_2019,meer_direct_2021,cogulu_direct_2021}, whereas our results show that domains in a-plane $\alpha$-Fe$_2$O$_3$ can be readily imaged by a table-top technique.

It is theoretically suggested that the damping-like SOT, not the field-like one, is needed to switch the N\'eel vector~\cite{baltz_antiferromagnetic_2018}.
The magnitude of the damping-like SOT measured in our work is considerably larger compared to previous reports on $\alpha$-Fe$_2$O$_3$/HM bilayers~\cite{cogulu_quantifying_2022,zhang_control_2022}.
This is great news for the efficient electrical control of the AFM order, but it also raises a question about the origin of the enhanced torque.
It was proposed that a low spin-mixing conductance of the thin film $\alpha$-Fe$_2$O$_3$/HM interface can explain a small magnitude of the damping-like SOT~\cite{cogulu_quantifying_2022,zhang_control_2022}.
Consequently, a high spin-mixing conductance of the a-plane terminated $\alpha$-Fe$_2$O$_3$ interface could account for the enhanced DL-SOT in our samples.
Contrary to these works, a spin-pumping study on bulk c-plane $\alpha$-Fe$_2$O$_3$/HM samples~\cite{wang_spin_2021} estimated the spin-mixing conductance to be two orders of magnitude larger than those reported in other FM/HM and AFM/FM systems.
This indicates that the quality of the thin films and bulk crystals could differ significantly.
We believe that more studies are needed to systematically investigate the interface between $\alpha$-Fe$_2$O$_3$ and a heavy metal, 
%of different orientations and quality
in order to understand and maximize the SOT in this promising material system.

In conclusion, we utilized current-modulated MOKE to characterize spin-orbit torque in a-plane $\alpha$-Fe$_{2}$O$_{3}$/Pt bilayers.
We found that damping-like SOT is two orders of magnitude larger than reported in c-plane and r-plane samples.
Our study identifies a-plane $\alpha$-Fe$_2$O$_3$ as a promising candidate to realize efficient SOT switching and calls for the development of thin films with a-plane orientation.

%For $|H_x| > |H_{sf}|$ and a small effective damping-like field $H_{DL} \ll |H_x - H_{sf}|$ the out-of-plane current-induced magnetization can be written as:
%\begin{equation}
%\dfrac{\Delta m^{net}_z}{m_{net}} = \dfrac{\Delta \varepsilon_K^j}{\varepsilon_K^{max}} = \dfrac{\mu_0 H_{DL}}{|\mu_0 H_x - \mu_0 H_{sf}|}
% \label{eq:KE_normalization}
%\end{equation}

%\section*{Acknowledgements}
\vspace{4pt}
This research was primarily supported (I.L., D.R., F.Y., R.K.K.) by the Center for Emergent Materials, an NSF MRSEC, under award number DMR-2011876 and partially supported (J.M., F.Y.) by the Department of Energy, Office of Science, Basic Energy Sciences, under Grant No. DE-SC0001304. R.C. and H.Z. were supported by the Air Force Office of Scientific Research under Grant No. FA9550-19-1-0307.

\bibliography{hematite.bib}

\begin{thebibliography}{32}%
\makeatletter
\providecommand \@ifxundefined [1]{%
 \@ifx{#1\undefined}
}%
\providecommand \@ifnum [1]{%
 \ifnum #1\expandafter \@firstoftwo
 \else \expandafter \@secondoftwo
 \fi
}%
\providecommand \@ifx [1]{%
 \ifx #1\expandafter \@firstoftwo
 \else \expandafter \@secondoftwo
 \fi
}%
\providecommand \natexlab [1]{#1}%
\providecommand \enquote  [1]{``#1''}%
\providecommand \bibnamefont  [1]{#1}%
\providecommand \bibfnamefont [1]{#1}%
\providecommand \citenamefont [1]{#1}%
\providecommand \href@noop [0]{\@secondoftwo}%
\providecommand \href [0]{\begingroup \@sanitize@url \@href}%
\providecommand \@href[1]{\@@startlink{#1}\@@href}%
\providecommand \@@href[1]{\endgroup#1\@@endlink}%
\providecommand \@sanitize@url [0]{\catcode `\\12\catcode `\$12\catcode `\&12\catcode `\#12\catcode `\^12\catcode `\_12\catcode `\%12\relax}%
\providecommand \@@startlink[1]{}%
\providecommand \@@endlink[0]{}%
\providecommand \url  [0]{\begingroup\@sanitize@url \@url }%
\providecommand \@url [1]{\endgroup\@href {#1}{\urlprefix }}%
\providecommand \urlprefix  [0]{URL }%
\providecommand \Eprint [0]{\href }%
\providecommand \doibase [0]{https://doi.org/}%
\providecommand \selectlanguage [0]{\@gobble}%
\providecommand \bibinfo  [0]{\@secondoftwo}%
\providecommand \bibfield  [0]{\@secondoftwo}%
\providecommand \translation [1]{[#1]}%
\providecommand \BibitemOpen [0]{}%
\providecommand \bibitemStop [0]{}%
\providecommand \bibitemNoStop [0]{.\EOS\space}%
\providecommand \EOS [0]{\spacefactor3000\relax}%
\providecommand \BibitemShut  [1]{\csname bibitem#1\endcsname}%
\let\auto@bib@innerbib\@empty
%</preamble>
\bibitem [{\citenamefont {Baltz}\ \emph {et~al.}(2018)\citenamefont {Baltz}, \citenamefont {Manchon}, \citenamefont {Tsoi}, \citenamefont {Moriyama}, \citenamefont {Ono},\ and\ \citenamefont {Tserkovnyak}}]{baltz_antiferromagnetic_2018}%
  \BibitemOpen
  \bibfield  {author} {\bibinfo {author} {\bibfnamefont {V.}~\bibnamefont {Baltz}}, \bibinfo {author} {\bibfnamefont {A.}~\bibnamefont {Manchon}}, \bibinfo {author} {\bibfnamefont {M.}~\bibnamefont {Tsoi}}, \bibinfo {author} {\bibfnamefont {T.}~\bibnamefont {Moriyama}}, \bibinfo {author} {\bibfnamefont {T.}~\bibnamefont {Ono}},\ and\ \bibinfo {author} {\bibfnamefont {Y.}~\bibnamefont {Tserkovnyak}},\ }\bibfield  {title} {\bibinfo {title} {Antiferromagnetic spintronics},\ }\href {https://doi.org/10.1103/RevModPhys.90.015005} {\bibfield  {journal} {\bibinfo  {journal} {Rev. Mod. Phys.}\ }\textbf {\bibinfo {volume} {90}},\ \bibinfo {pages} {015005} (\bibinfo {year} {2018})}\BibitemShut {NoStop}%
\bibitem [{\citenamefont {Fukami}\ \emph {et~al.}(2020)\citenamefont {Fukami}, \citenamefont {Lorenz},\ and\ \citenamefont {Gomonay}}]{fukami_antiferromagnetic_2020}%
  \BibitemOpen
  \bibfield  {author} {\bibinfo {author} {\bibfnamefont {S.}~\bibnamefont {Fukami}}, \bibinfo {author} {\bibfnamefont {V.~O.}\ \bibnamefont {Lorenz}},\ and\ \bibinfo {author} {\bibfnamefont {O.}~\bibnamefont {Gomonay}},\ }\bibfield  {title} {\bibinfo {title} {Antiferromagnetic spintronics},\ }\href {https://doi.org/10.1063/5.0023614} {\bibfield  {journal} {\bibinfo  {journal} {Journal of Applied Physics}\ }\textbf {\bibinfo {volume} {128}},\ \bibinfo {pages} {070401} (\bibinfo {year} {2020})}\BibitemShut {NoStop}%
\bibitem [{\citenamefont {Han}\ \emph {et~al.}(2023)\citenamefont {Han}, \citenamefont {Cheng}, \citenamefont {Liu}, \citenamefont {Ohno},\ and\ \citenamefont {Fukami}}]{han_coherent_2023}%
  \BibitemOpen
  \bibfield  {author} {\bibinfo {author} {\bibfnamefont {J.}~\bibnamefont {Han}}, \bibinfo {author} {\bibfnamefont {R.}~\bibnamefont {Cheng}}, \bibinfo {author} {\bibfnamefont {L.}~\bibnamefont {Liu}}, \bibinfo {author} {\bibfnamefont {H.}~\bibnamefont {Ohno}},\ and\ \bibinfo {author} {\bibfnamefont {S.}~\bibnamefont {Fukami}},\ }\bibfield  {title} {\bibinfo {title} {Coherent antiferromagnetic spintronics},\ }\href {https://doi.org/10.1038/s41563-023-01492-6} {\bibfield  {journal} {\bibinfo  {journal} {Nat. Mater.}\ }\textbf {\bibinfo {volume} {22}},\ \bibinfo {pages} {684} (\bibinfo {year} {2023})}\BibitemShut {NoStop}%
\bibitem [{\citenamefont {Liu}\ \emph {et~al.}(2011)\citenamefont {Liu}, \citenamefont {Moriyama}, \citenamefont {Ralph},\ and\ \citenamefont {Buhrman}}]{liu_spin-torque_2011}%
  \BibitemOpen
  \bibfield  {author} {\bibinfo {author} {\bibfnamefont {L.}~\bibnamefont {Liu}}, \bibinfo {author} {\bibfnamefont {T.}~\bibnamefont {Moriyama}}, \bibinfo {author} {\bibfnamefont {D.~C.}\ \bibnamefont {Ralph}},\ and\ \bibinfo {author} {\bibfnamefont {R.~A.}\ \bibnamefont {Buhrman}},\ }\bibfield  {title} {\bibinfo {title} {Spin-{Torque} {Ferromagnetic} {Resonance} {Induced} by the {Spin} {Hall} {Effect}},\ }\href {https://doi.org/10.1103/PhysRevLett.106.036601} {\bibfield  {journal} {\bibinfo  {journal} {Phys. Rev. Lett.}\ }\textbf {\bibinfo {volume} {106}},\ \bibinfo {pages} {036601} (\bibinfo {year} {2011})}\BibitemShut {NoStop}%
\bibitem [{\citenamefont {Miron}\ \emph {et~al.}(2011)\citenamefont {Miron}, \citenamefont {Garello}, \citenamefont {Gaudin}, \citenamefont {Zermatten}, \citenamefont {Costache}, \citenamefont {Auffret}, \citenamefont {Bandiera}, \citenamefont {Rodmacq}, \citenamefont {Schuhl},\ and\ \citenamefont {Gambardella}}]{miron_perpendicular_2011}%
  \BibitemOpen
  \bibfield  {author} {\bibinfo {author} {\bibfnamefont {I.~M.}\ \bibnamefont {Miron}}, \bibinfo {author} {\bibfnamefont {K.}~\bibnamefont {Garello}}, \bibinfo {author} {\bibfnamefont {G.}~\bibnamefont {Gaudin}}, \bibinfo {author} {\bibfnamefont {P.-J.}\ \bibnamefont {Zermatten}}, \bibinfo {author} {\bibfnamefont {M.~V.}\ \bibnamefont {Costache}}, \bibinfo {author} {\bibfnamefont {S.}~\bibnamefont {Auffret}}, \bibinfo {author} {\bibfnamefont {S.}~\bibnamefont {Bandiera}}, \bibinfo {author} {\bibfnamefont {B.}~\bibnamefont {Rodmacq}}, \bibinfo {author} {\bibfnamefont {A.}~\bibnamefont {Schuhl}},\ and\ \bibinfo {author} {\bibfnamefont {P.}~\bibnamefont {Gambardella}},\ }\bibfield  {title} {\bibinfo {title} {Perpendicular switching of a single ferromagnetic layer induced by in-plane current injection},\ }\href {https://doi.org/10.1038/nature10309} {\bibfield  {journal} {\bibinfo  {journal} {Nature}\ }\textbf {\bibinfo {volume} {476}},\ \bibinfo {pages} {189} (\bibinfo {year} {2011})}\BibitemShut {NoStop}%
\bibitem [{\citenamefont {Liu}\ \emph {et~al.}(2012)\citenamefont {Liu}, \citenamefont {Pai}, \citenamefont {Li}, \citenamefont {Tseng}, \citenamefont {Ralph},\ and\ \citenamefont {Buhrman}}]{liu_spin-torque_2012}%
  \BibitemOpen
  \bibfield  {author} {\bibinfo {author} {\bibfnamefont {L.}~\bibnamefont {Liu}}, \bibinfo {author} {\bibfnamefont {C.-F.}\ \bibnamefont {Pai}}, \bibinfo {author} {\bibfnamefont {Y.}~\bibnamefont {Li}}, \bibinfo {author} {\bibfnamefont {H.~W.}\ \bibnamefont {Tseng}}, \bibinfo {author} {\bibfnamefont {D.~C.}\ \bibnamefont {Ralph}},\ and\ \bibinfo {author} {\bibfnamefont {R.~A.}\ \bibnamefont {Buhrman}},\ }\bibfield  {title} {\bibinfo {title} {Spin-{Torque} {Switching} with the {Giant} {Spin} {Hall} {Effect} of {Tantalum}},\ }\href {https://doi.org/10.1126/science.1218197} {\bibfield  {journal} {\bibinfo  {journal} {Science}\ }\textbf {\bibinfo {volume} {336}},\ \bibinfo {pages} {555} (\bibinfo {year} {2012})}\BibitemShut {NoStop}%
\bibitem [{\citenamefont {Manchon}\ \emph {et~al.}(2019)\citenamefont {Manchon}, \citenamefont {Železný}, \citenamefont {Miron}, \citenamefont {Jungwirth}, \citenamefont {Sinova}, \citenamefont {Thiaville}, \citenamefont {Garello},\ and\ \citenamefont {Gambardella}}]{manchon_current-induced_2019}%
  \BibitemOpen
  \bibfield  {author} {\bibinfo {author} {\bibfnamefont {A.}~\bibnamefont {Manchon}}, \bibinfo {author} {\bibfnamefont {J.}~\bibnamefont {Železný}}, \bibinfo {author} {\bibfnamefont {I.}~\bibnamefont {Miron}}, \bibinfo {author} {\bibfnamefont {T.}~\bibnamefont {Jungwirth}}, \bibinfo {author} {\bibfnamefont {J.}~\bibnamefont {Sinova}}, \bibinfo {author} {\bibfnamefont {A.}~\bibnamefont {Thiaville}}, \bibinfo {author} {\bibfnamefont {K.}~\bibnamefont {Garello}},\ and\ \bibinfo {author} {\bibfnamefont {P.}~\bibnamefont {Gambardella}},\ }\bibfield  {title} {\bibinfo {title} {Current-induced spin-orbit torques in ferromagnetic and antiferromagnetic systems},\ }\href {https://doi.org/10.1103/RevModPhys.91.035004} {\bibfield  {journal} {\bibinfo  {journal} {Rev. Mod. Phys.}\ }\textbf {\bibinfo {volume} {91}},\ \bibinfo {pages} {035004} (\bibinfo {year} {2019})}\BibitemShut {NoStop}%
\bibitem [{\citenamefont {Chiang}\ \emph {et~al.}(2019)\citenamefont {Chiang}, \citenamefont {Huang}, \citenamefont {Qu}, \citenamefont {Wu},\ and\ \citenamefont {Chien}}]{chiang_absence_2019}%
  \BibitemOpen
  \bibfield  {author} {\bibinfo {author} {\bibfnamefont {C.}~\bibnamefont {Chiang}}, \bibinfo {author} {\bibfnamefont {S.}~\bibnamefont {Huang}}, \bibinfo {author} {\bibfnamefont {D.}~\bibnamefont {Qu}}, \bibinfo {author} {\bibfnamefont {P.}~\bibnamefont {Wu}},\ and\ \bibinfo {author} {\bibfnamefont {C.}~\bibnamefont {Chien}},\ }\bibfield  {title} {\bibinfo {title} {Absence of {Evidence} of {Electrical} {Switching} of the {Antiferromagnetic} {N\'eel} {Vector}},\ }\href {https://doi.org/10.1103/PhysRevLett.123.227203} {\bibfield  {journal} {\bibinfo  {journal} {Phys. Rev. Lett.}\ }\textbf {\bibinfo {volume} {123}},\ \bibinfo {pages} {227203} (\bibinfo {year} {2019})}\BibitemShut {NoStop}%
\bibitem [{\citenamefont {Baldrati}\ \emph {et~al.}(2019)\citenamefont {Baldrati}, \citenamefont {Gomonay}, \citenamefont {Ross}, \citenamefont {Filianina}, \citenamefont {Lebrun}, \citenamefont {Ramos}, \citenamefont {Leveille}, \citenamefont {Fuhrmann}, \citenamefont {Forrest}, \citenamefont {Maccherozzi}, \citenamefont {Valencia}, \citenamefont {Kronast}, \citenamefont {Saitoh}, \citenamefont {Sinova},\ and\ \citenamefont {Kläui}}]{baldrati_mechanism_2019}%
  \BibitemOpen
  \bibfield  {author} {\bibinfo {author} {\bibfnamefont {L.}~\bibnamefont {Baldrati}}, \bibinfo {author} {\bibfnamefont {O.}~\bibnamefont {Gomonay}}, \bibinfo {author} {\bibfnamefont {A.}~\bibnamefont {Ross}}, \bibinfo {author} {\bibfnamefont {M.}~\bibnamefont {Filianina}}, \bibinfo {author} {\bibfnamefont {R.}~\bibnamefont {Lebrun}}, \bibinfo {author} {\bibfnamefont {R.}~\bibnamefont {Ramos}}, \bibinfo {author} {\bibfnamefont {C.}~\bibnamefont {Leveille}}, \bibinfo {author} {\bibfnamefont {F.}~\bibnamefont {Fuhrmann}}, \bibinfo {author} {\bibfnamefont {T.~R.}\ \bibnamefont {Forrest}}, \bibinfo {author} {\bibfnamefont {F.}~\bibnamefont {Maccherozzi}}, \bibinfo {author} {\bibfnamefont {S.}~\bibnamefont {Valencia}}, \bibinfo {author} {\bibfnamefont {F.}~\bibnamefont {Kronast}}, \bibinfo {author} {\bibfnamefont {E.}~\bibnamefont {Saitoh}}, \bibinfo {author} {\bibfnamefont {J.}~\bibnamefont {Sinova}},\ and\ \bibinfo {author} {\bibfnamefont {M.}~\bibnamefont {Kläui}},\ }\bibfield  {title} {\bibinfo {title}
  {Mechanism of {N}{\textbackslash}'eel {Order} {Switching} in {Antiferromagnetic} {Thin} {Films} {Revealed} by {Magnetotransport} and {Direct} {Imaging}},\ }\href {https://doi.org/10.1103/PhysRevLett.123.177201} {\bibfield  {journal} {\bibinfo  {journal} {Phys. Rev. Lett.}\ }\textbf {\bibinfo {volume} {123}},\ \bibinfo {pages} {177201} (\bibinfo {year} {2019})}\BibitemShut {NoStop}%
\bibitem [{\citenamefont {Zhang}\ \emph {et~al.}(2019)\citenamefont {Zhang}, \citenamefont {Finley}, \citenamefont {Safi},\ and\ \citenamefont {Liu}}]{zhang_quantitative_2019}%
  \BibitemOpen
  \bibfield  {author} {\bibinfo {author} {\bibfnamefont {P.}~\bibnamefont {Zhang}}, \bibinfo {author} {\bibfnamefont {J.}~\bibnamefont {Finley}}, \bibinfo {author} {\bibfnamefont {T.}~\bibnamefont {Safi}},\ and\ \bibinfo {author} {\bibfnamefont {L.}~\bibnamefont {Liu}},\ }\bibfield  {title} {\bibinfo {title} {Quantitative {Study} on {Current}-{Induced} {Effect} in an {Antiferromagnet} {Insulator}/{Pt} {Bilayer} {Film}},\ }\href {https://doi.org/10.1103/PhysRevLett.123.247206} {\bibfield  {journal} {\bibinfo  {journal} {Phys. Rev. Lett.}\ }\textbf {\bibinfo {volume} {123}},\ \bibinfo {pages} {247206} (\bibinfo {year} {2019})}\BibitemShut {NoStop}%
\bibitem [{\citenamefont {Churikova}\ \emph {et~al.}(2020)\citenamefont {Churikova}, \citenamefont {Bono}, \citenamefont {Neltner}, \citenamefont {Wittmann}, \citenamefont {Scipioni}, \citenamefont {Shepard}, \citenamefont {Newhouse-Illige}, \citenamefont {Greer},\ and\ \citenamefont {Beach}}]{churikova_non-magnetic_2020}%
  \BibitemOpen
  \bibfield  {author} {\bibinfo {author} {\bibfnamefont {A.}~\bibnamefont {Churikova}}, \bibinfo {author} {\bibfnamefont {D.}~\bibnamefont {Bono}}, \bibinfo {author} {\bibfnamefont {B.}~\bibnamefont {Neltner}}, \bibinfo {author} {\bibfnamefont {A.}~\bibnamefont {Wittmann}}, \bibinfo {author} {\bibfnamefont {L.}~\bibnamefont {Scipioni}}, \bibinfo {author} {\bibfnamefont {A.}~\bibnamefont {Shepard}}, \bibinfo {author} {\bibfnamefont {T.}~\bibnamefont {Newhouse-Illige}}, \bibinfo {author} {\bibfnamefont {J.}~\bibnamefont {Greer}},\ and\ \bibinfo {author} {\bibfnamefont {G.~S.~D.}\ \bibnamefont {Beach}},\ }\bibfield  {title} {\bibinfo {title} {Non-magnetic origin of spin {Hall} magnetoresistance-like signals in {Pt} films and epitaxial {NiO}/{Pt} bilayers},\ }\href {https://doi.org/10.1063/1.5134814} {\bibfield  {journal} {\bibinfo  {journal} {Applied Physics Letters}\ }\textbf {\bibinfo {volume} {116}},\ \bibinfo {pages} {022410} (\bibinfo {year} {2020})}\BibitemShut {NoStop}%
\bibitem [{\citenamefont {Meer}\ \emph {et~al.}(2021)\citenamefont {Meer}, \citenamefont {Schreiber}, \citenamefont {Schmitt}, \citenamefont {Ramos}, \citenamefont {Saitoh}, \citenamefont {Gomonay}, \citenamefont {Sinova}, \citenamefont {Baldrati},\ and\ \citenamefont {Kläui}}]{meer_direct_2021}%
  \BibitemOpen
  \bibfield  {author} {\bibinfo {author} {\bibfnamefont {H.}~\bibnamefont {Meer}}, \bibinfo {author} {\bibfnamefont {F.}~\bibnamefont {Schreiber}}, \bibinfo {author} {\bibfnamefont {C.}~\bibnamefont {Schmitt}}, \bibinfo {author} {\bibfnamefont {R.}~\bibnamefont {Ramos}}, \bibinfo {author} {\bibfnamefont {E.}~\bibnamefont {Saitoh}}, \bibinfo {author} {\bibfnamefont {O.}~\bibnamefont {Gomonay}}, \bibinfo {author} {\bibfnamefont {J.}~\bibnamefont {Sinova}}, \bibinfo {author} {\bibfnamefont {L.}~\bibnamefont {Baldrati}},\ and\ \bibinfo {author} {\bibfnamefont {M.}~\bibnamefont {Kläui}},\ }\bibfield  {title} {\bibinfo {title} {Direct {Imaging} of {Current}-{Induced} {Antiferromagnetic} {Switching} {Revealing} a {Pure} {Thermomagnetoelastic} {Switching} {Mechanism} in {NiO}},\ }\href {https://doi.org/10.1021/acs.nanolett.0c03367} {\bibfield  {journal} {\bibinfo  {journal} {Nano Lett.}\ }\textbf {\bibinfo {volume} {21}},\ \bibinfo {pages} {114} (\bibinfo {year} {2021})}\BibitemShut {NoStop}%
\bibitem [{\citenamefont {Fan}\ \emph {et~al.}(2014)\citenamefont {Fan}, \citenamefont {Celik}, \citenamefont {Wu}, \citenamefont {Ni}, \citenamefont {Lee}, \citenamefont {Lorenz},\ and\ \citenamefont {Xiao}}]{fan_quantifying_2014}%
  \BibitemOpen
  \bibfield  {author} {\bibinfo {author} {\bibfnamefont {X.}~\bibnamefont {Fan}}, \bibinfo {author} {\bibfnamefont {H.}~\bibnamefont {Celik}}, \bibinfo {author} {\bibfnamefont {J.}~\bibnamefont {Wu}}, \bibinfo {author} {\bibfnamefont {C.}~\bibnamefont {Ni}}, \bibinfo {author} {\bibfnamefont {K.-J.}\ \bibnamefont {Lee}}, \bibinfo {author} {\bibfnamefont {V.~O.}\ \bibnamefont {Lorenz}},\ and\ \bibinfo {author} {\bibfnamefont {J.~Q.}\ \bibnamefont {Xiao}},\ }\bibfield  {title} {\bibinfo {title} {Quantifying interface and bulk contributions to spin–orbit torque in magnetic bilayers},\ }\href {https://doi.org/10.1038/ncomms4042} {\bibfield  {journal} {\bibinfo  {journal} {Nat Commun}\ }\textbf {\bibinfo {volume} {5}},\ \bibinfo {pages} {3042} (\bibinfo {year} {2014})}\BibitemShut {NoStop}%
\bibitem [{\citenamefont {Fan}\ \emph {et~al.}(2016)\citenamefont {Fan}, \citenamefont {Mellnik}, \citenamefont {Wang}, \citenamefont {Reynolds}, \citenamefont {Wang}, \citenamefont {Celik}, \citenamefont {Lorenz}, \citenamefont {Ralph},\ and\ \citenamefont {Xiao}}]{fan_all-optical_2016}%
  \BibitemOpen
  \bibfield  {author} {\bibinfo {author} {\bibfnamefont {X.}~\bibnamefont {Fan}}, \bibinfo {author} {\bibfnamefont {A.~R.}\ \bibnamefont {Mellnik}}, \bibinfo {author} {\bibfnamefont {W.}~\bibnamefont {Wang}}, \bibinfo {author} {\bibfnamefont {N.}~\bibnamefont {Reynolds}}, \bibinfo {author} {\bibfnamefont {T.}~\bibnamefont {Wang}}, \bibinfo {author} {\bibfnamefont {H.}~\bibnamefont {Celik}}, \bibinfo {author} {\bibfnamefont {V.~O.}\ \bibnamefont {Lorenz}}, \bibinfo {author} {\bibfnamefont {D.~C.}\ \bibnamefont {Ralph}},\ and\ \bibinfo {author} {\bibfnamefont {J.~Q.}\ \bibnamefont {Xiao}},\ }\bibfield  {title} {\bibinfo {title} {All-optical vector measurement of spin-orbit-induced torques using both polar and quadratic magneto-optic {Kerr} effects},\ }\href {https://doi.org/10.1063/1.4962402} {\bibfield  {journal} {\bibinfo  {journal} {Applied Physics Letters}\ }\textbf {\bibinfo {volume} {109}},\ \bibinfo {pages} {122406} (\bibinfo {year} {2016})}\BibitemShut {NoStop}%
\bibitem [{\citenamefont {Wang}\ \emph {et~al.}(2019)\citenamefont {Wang}, \citenamefont {Wang}, \citenamefont {Amin}, \citenamefont {Wang}, \citenamefont {Radhakrishnan}, \citenamefont {Davidson}, \citenamefont {Allen}, \citenamefont {Silva}, \citenamefont {Ohldag}, \citenamefont {Balzar}, \citenamefont {Zink}, \citenamefont {Haney}, \citenamefont {Xiao}, \citenamefont {Cahill}, \citenamefont {Lorenz},\ and\ \citenamefont {Fan}}]{wang_anomalous_2019}%
  \BibitemOpen
  \bibfield  {author} {\bibinfo {author} {\bibfnamefont {W.}~\bibnamefont {Wang}}, \bibinfo {author} {\bibfnamefont {T.}~\bibnamefont {Wang}}, \bibinfo {author} {\bibfnamefont {V.~P.}\ \bibnamefont {Amin}}, \bibinfo {author} {\bibfnamefont {Y.}~\bibnamefont {Wang}}, \bibinfo {author} {\bibfnamefont {A.}~\bibnamefont {Radhakrishnan}}, \bibinfo {author} {\bibfnamefont {A.}~\bibnamefont {Davidson}}, \bibinfo {author} {\bibfnamefont {S.~R.}\ \bibnamefont {Allen}}, \bibinfo {author} {\bibfnamefont {T.~J.}\ \bibnamefont {Silva}}, \bibinfo {author} {\bibfnamefont {H.}~\bibnamefont {Ohldag}}, \bibinfo {author} {\bibfnamefont {D.}~\bibnamefont {Balzar}}, \bibinfo {author} {\bibfnamefont {B.~L.}\ \bibnamefont {Zink}}, \bibinfo {author} {\bibfnamefont {P.~M.}\ \bibnamefont {Haney}}, \bibinfo {author} {\bibfnamefont {J.~Q.}\ \bibnamefont {Xiao}}, \bibinfo {author} {\bibfnamefont {D.~G.}\ \bibnamefont {Cahill}}, \bibinfo {author} {\bibfnamefont {V.~O.}\ \bibnamefont {Lorenz}},\ and\ \bibinfo {author} {\bibfnamefont
  {X.}~\bibnamefont {Fan}},\ }\bibfield  {title} {\bibinfo {title} {Anomalous spin–orbit torques in magnetic single-layer films},\ }\href {https://doi.org/10.1038/s41565-019-0504-0} {\bibfield  {journal} {\bibinfo  {journal} {Nat. Nanotechnol.}\ }\textbf {\bibinfo {volume} {14}},\ \bibinfo {pages} {819} (\bibinfo {year} {2019})}\BibitemShut {NoStop}%
\bibitem [{\citenamefont {Lyalin}\ \emph {et~al.}(2021)\citenamefont {Lyalin}, \citenamefont {Cheng},\ and\ \citenamefont {Kawakami}}]{lyalin_spinorbit_2021}%
  \BibitemOpen
  \bibfield  {author} {\bibinfo {author} {\bibfnamefont {I.}~\bibnamefont {Lyalin}}, \bibinfo {author} {\bibfnamefont {S.}~\bibnamefont {Cheng}},\ and\ \bibinfo {author} {\bibfnamefont {R.~K.}\ \bibnamefont {Kawakami}},\ }\bibfield  {title} {\bibinfo {title} {Spin–{Orbit} {Torque} in {Bilayers} of {Kagome} {Ferromagnet} {Fe$_3$Sn$_2$} and {Pt}},\ }\href {https://doi.org/10.1021/acs.nanolett.1c02270} {\bibfield  {journal} {\bibinfo  {journal} {Nano Lett.}\ }\textbf {\bibinfo {volume} {21}},\ \bibinfo {pages} {6975} (\bibinfo {year} {2021})}\BibitemShut {NoStop}%
\bibitem [{\citenamefont {Pham}\ \emph {et~al.}(2023)\citenamefont {Pham}, \citenamefont {Ko},\ and\ \citenamefont {Choi}}]{pham_ferromagnetic_2023}%
  \BibitemOpen
  \bibfield  {author} {\bibinfo {author} {\bibfnamefont {N.~L.~L.}\ \bibnamefont {Pham}}, \bibinfo {author} {\bibfnamefont {K.-H.}\ \bibnamefont {Ko}},\ and\ \bibinfo {author} {\bibfnamefont {G.-M.}\ \bibnamefont {Choi}},\ }\bibfield  {title} {\bibinfo {title} {Ferromagnetic material dependence of spin–orbit torque in {PtMn}/ferromagnet bilayer},\ }\href {https://doi.org/10.1063/5.0166503} {\bibfield  {journal} {\bibinfo  {journal} {Applied Physics Letters}\ }\textbf {\bibinfo {volume} {123}},\ \bibinfo {pages} {162402} (\bibinfo {year} {2023})}\BibitemShut {NoStop}%
\bibitem [{\citenamefont {Cogulu}\ \emph {et~al.}(2022)\citenamefont {Cogulu}, \citenamefont {Zhang}, \citenamefont {Statuto}, \citenamefont {Cheng}, \citenamefont {Yang}, \citenamefont {Cheng},\ and\ \citenamefont {Kent}}]{cogulu_quantifying_2022}%
  \BibitemOpen
  \bibfield  {author} {\bibinfo {author} {\bibfnamefont {E.}~\bibnamefont {Cogulu}}, \bibinfo {author} {\bibfnamefont {H.}~\bibnamefont {Zhang}}, \bibinfo {author} {\bibfnamefont {N.~N.}\ \bibnamefont {Statuto}}, \bibinfo {author} {\bibfnamefont {Y.}~\bibnamefont {Cheng}}, \bibinfo {author} {\bibfnamefont {F.}~\bibnamefont {Yang}}, \bibinfo {author} {\bibfnamefont {R.}~\bibnamefont {Cheng}},\ and\ \bibinfo {author} {\bibfnamefont {A.~D.}\ \bibnamefont {Kent}},\ }\bibfield  {title} {\bibinfo {title} {Quantifying {Spin}-{Orbit} {Torques} in {Antiferromagnet}–{Heavy}-{Metal} {Heterostructures}},\ }\href {https://doi.org/10.1103/PhysRevLett.128.247204} {\bibfield  {journal} {\bibinfo  {journal} {Phys. Rev. Lett.}\ }\textbf {\bibinfo {volume} {128}},\ \bibinfo {pages} {247204} (\bibinfo {year} {2022})}\BibitemShut {NoStop}%
\bibitem [{\citenamefont {Zhang}\ \emph {et~al.}(2022)\citenamefont {Zhang}, \citenamefont {Chou}, \citenamefont {Yun}, \citenamefont {McGoldrick}, \citenamefont {Hou}, \citenamefont {Mkhoyan},\ and\ \citenamefont {Liu}}]{zhang_control_2022}%
  \BibitemOpen
  \bibfield  {author} {\bibinfo {author} {\bibfnamefont {P.}~\bibnamefont {Zhang}}, \bibinfo {author} {\bibfnamefont {C.-T.}\ \bibnamefont {Chou}}, \bibinfo {author} {\bibfnamefont {H.}~\bibnamefont {Yun}}, \bibinfo {author} {\bibfnamefont {B.}~\bibnamefont {McGoldrick}}, \bibinfo {author} {\bibfnamefont {J.}~\bibnamefont {Hou}}, \bibinfo {author} {\bibfnamefont {K.~A.}\ \bibnamefont {Mkhoyan}},\ and\ \bibinfo {author} {\bibfnamefont {L.}~\bibnamefont {Liu}},\ }\bibfield  {title} {\bibinfo {title} {Control of {Néel} {Vector} with {Spin}-{Orbit} {Torques} in an {Antiferromagnetic} {Insulator} with {Tilted} {Easy} {Plane}},\ }\href {https://doi.org/10.1103/PhysRevLett.129.017203} {\bibfield  {journal} {\bibinfo  {journal} {Phys. Rev. Lett.}\ }\textbf {\bibinfo {volume} {129}},\ \bibinfo {pages} {017203} (\bibinfo {year} {2022})}\BibitemShut {NoStop}%
\bibitem [{\citenamefont {Ahmed}\ \emph {et~al.}(2019)\citenamefont {Ahmed}, \citenamefont {Lee}, \citenamefont {Bagués}, \citenamefont {McCullian}, \citenamefont {Thabt}, \citenamefont {Perrine}, \citenamefont {Wu}, \citenamefont {Rowland}, \citenamefont {Randeria}, \citenamefont {Hammel}, \citenamefont {McComb},\ and\ \citenamefont {Yang}}]{ahmed_spin-hall_2019}%
  \BibitemOpen
  \bibfield  {author} {\bibinfo {author} {\bibfnamefont {A.~S.}\ \bibnamefont {Ahmed}}, \bibinfo {author} {\bibfnamefont {A.~J.}\ \bibnamefont {Lee}}, \bibinfo {author} {\bibfnamefont {N.}~\bibnamefont {Bagués}}, \bibinfo {author} {\bibfnamefont {B.~A.}\ \bibnamefont {McCullian}}, \bibinfo {author} {\bibfnamefont {A.~M.~A.}\ \bibnamefont {Thabt}}, \bibinfo {author} {\bibfnamefont {A.}~\bibnamefont {Perrine}}, \bibinfo {author} {\bibfnamefont {P.-K.}\ \bibnamefont {Wu}}, \bibinfo {author} {\bibfnamefont {J.~R.}\ \bibnamefont {Rowland}}, \bibinfo {author} {\bibfnamefont {M.}~\bibnamefont {Randeria}}, \bibinfo {author} {\bibfnamefont {P.~C.}\ \bibnamefont {Hammel}}, \bibinfo {author} {\bibfnamefont {D.~W.}\ \bibnamefont {McComb}},\ and\ \bibinfo {author} {\bibfnamefont {F.}~\bibnamefont {Yang}},\ }\bibfield  {title} {\bibinfo {title} {Spin-{Hall} {Topological} {Hall} {Effect} in {Highly} {Tunable} {Pt}/{Ferrimagnetic}-{Insulator} {Bilayers}},\ }\href {https://doi.org/10.1021/acs.nanolett.9b02265} {\bibfield
  {journal} {\bibinfo  {journal} {Nano Lett.}\ }\textbf {\bibinfo {volume} {19}},\ \bibinfo {pages} {5683} (\bibinfo {year} {2019})}\BibitemShut {NoStop}%
\bibitem [{\citenamefont {Avci}\ \emph {et~al.}(2017)\citenamefont {Avci}, \citenamefont {Quindeau}, \citenamefont {Pai}, \citenamefont {Mann}, \citenamefont {Caretta}, \citenamefont {Tang}, \citenamefont {Onbasli}, \citenamefont {Ross},\ and\ \citenamefont {Beach}}]{avci_current-induced_2017}%
  \BibitemOpen
  \bibfield  {author} {\bibinfo {author} {\bibfnamefont {C.~O.}\ \bibnamefont {Avci}}, \bibinfo {author} {\bibfnamefont {A.}~\bibnamefont {Quindeau}}, \bibinfo {author} {\bibfnamefont {C.-F.}\ \bibnamefont {Pai}}, \bibinfo {author} {\bibfnamefont {M.}~\bibnamefont {Mann}}, \bibinfo {author} {\bibfnamefont {L.}~\bibnamefont {Caretta}}, \bibinfo {author} {\bibfnamefont {A.~S.}\ \bibnamefont {Tang}}, \bibinfo {author} {\bibfnamefont {M.~C.}\ \bibnamefont {Onbasli}}, \bibinfo {author} {\bibfnamefont {C.~A.}\ \bibnamefont {Ross}},\ and\ \bibinfo {author} {\bibfnamefont {G.~S.~D.}\ \bibnamefont {Beach}},\ }\bibfield  {title} {\bibinfo {title} {Current-induced switching in a magnetic insulator},\ }\href {https://doi.org/10.1038/nmat4812} {\bibfield  {journal} {\bibinfo  {journal} {Nature Mater}\ }\textbf {\bibinfo {volume} {16}},\ \bibinfo {pages} {309} (\bibinfo {year} {2017})}\BibitemShut {NoStop}%
\bibitem [{\citenamefont {Li}\ \emph {et~al.}(2023)\citenamefont {Li}, \citenamefont {Liu}, \citenamefont {Li}, \citenamefont {Zhao}, \citenamefont {An},\ and\ \citenamefont {Ando}}]{li_giant_2023}%
  \BibitemOpen
  \bibfield  {author} {\bibinfo {author} {\bibfnamefont {T.}~\bibnamefont {Li}}, \bibinfo {author} {\bibfnamefont {L.}~\bibnamefont {Liu}}, \bibinfo {author} {\bibfnamefont {X.}~\bibnamefont {Li}}, \bibinfo {author} {\bibfnamefont {X.}~\bibnamefont {Zhao}}, \bibinfo {author} {\bibfnamefont {H.}~\bibnamefont {An}},\ and\ \bibinfo {author} {\bibfnamefont {K.}~\bibnamefont {Ando}},\ }\bibfield  {title} {\bibinfo {title} {Giant {Orbital}-to-{Spin} {Conversion} for {Efficient} {Current}-{Induced} {Magnetization} {Switching} of {Ferrimagnetic} {Insulator}},\ }\href {https://doi.org/10.1021/acs.nanolett.3c02104} {\bibfield  {journal} {\bibinfo  {journal} {Nano Lett.}\ }\textbf {\bibinfo {volume} {23}},\ \bibinfo {pages} {7174} (\bibinfo {year} {2023})}\BibitemShut {NoStop}%
\bibitem [{Sup()}]{SupplMat}%
  \BibitemOpen
  \href@noop {} {}\bibinfo {note} {See Supplemental Material for details on the additional magnetic circular dichroism and angular-dependent magnetoresistance measurements, current-modulated MOKE measurements of another $\alpha$-Fe$_2$O$_3$ device, and theoretical derivation of the fitting formula, which includes Refs.~\cite{zhang2022theory,SpinSuperfluidAFM}.}\BibitemShut {Stop}%
\bibitem [{\citenamefont {Williamson}\ and\ \citenamefont {Foner}(1964)}]{williamson_antiferromagnetic_1964}%
  \BibitemOpen
  \bibfield  {author} {\bibinfo {author} {\bibfnamefont {S.~J.}\ \bibnamefont {Williamson}}\ and\ \bibinfo {author} {\bibfnamefont {S.}~\bibnamefont {Foner}},\ }\bibfield  {title} {\bibinfo {title} {{Antiferromagnetic Resonance in Systems with Dzyaloshinsky-Moriya Coupling; Orientation Dependence in $\alpha$-Fe$_2$O$_3$}},\ }\href {https://doi.org/10.1103/PhysRev.136.A1102} {\bibfield  {journal} {\bibinfo  {journal} {Phys. Rev.}\ }\textbf {\bibinfo {volume} {136}},\ \bibinfo {pages} {A1102} (\bibinfo {year} {1964})}\BibitemShut {NoStop}%
\bibitem [{\citenamefont {Mizushima}\ and\ \citenamefont {Iida}(1966)}]{mizushima_effective_1966}%
  \BibitemOpen
  \bibfield  {author} {\bibinfo {author} {\bibfnamefont {K.}~\bibnamefont {Mizushima}}\ and\ \bibinfo {author} {\bibfnamefont {S.}~\bibnamefont {Iida}},\ }\bibfield  {title} {\bibinfo {title} {{Effective In-Plane Anisotropy Field in $\alpha$-Fe$_2$O$_3$}},\ }\href {https://doi.org/10.1143/JPSJ.21.1521} {\bibfield  {journal} {\bibinfo  {journal} {Journal of the Physical Society of Japan}\ }\textbf {\bibinfo {volume} {21}},\ \bibinfo {pages} {1521} (\bibinfo {year} {1966})}\BibitemShut {NoStop}%
\bibitem [{\citenamefont {Elliston}\ and\ \citenamefont {Troup}(1968)}]{elliston_some_1968}%
  \BibitemOpen
  \bibfield  {author} {\bibinfo {author} {\bibfnamefont {P.~R.}\ \bibnamefont {Elliston}}\ and\ \bibinfo {author} {\bibfnamefont {G.~J.}\ \bibnamefont {Troup}},\ }\bibfield  {title} {\bibinfo {title} {{Some antiferromagnetic resonance measurements in $\alpha$-Fe$_2$O$_3$}},\ }\href {https://api.semanticscholar.org/CorpusID:119601276} {\bibfield  {journal} {\bibinfo  {journal} {Journal of Physics C: Solid State Physics}\ }\textbf {\bibinfo {volume} {1}},\ \bibinfo {pages} {169} (\bibinfo {year} {1968})}\BibitemShut {NoStop}%
\bibitem [{\citenamefont {Lebrun}\ \emph {et~al.}(2019)\citenamefont {Lebrun}, \citenamefont {Ross}, \citenamefont {Gomonay}, \citenamefont {Bender}, \citenamefont {Baldrati}, \citenamefont {Kronast}, \citenamefont {Qaiumzadeh}, \citenamefont {Sinova}, \citenamefont {Brataas}, \citenamefont {Duine},\ and\ \citenamefont {Kläui}}]{lebrun_anisotropies_2019}%
  \BibitemOpen
  \bibfield  {author} {\bibinfo {author} {\bibfnamefont {R.}~\bibnamefont {Lebrun}}, \bibinfo {author} {\bibfnamefont {A.}~\bibnamefont {Ross}}, \bibinfo {author} {\bibfnamefont {O.}~\bibnamefont {Gomonay}}, \bibinfo {author} {\bibfnamefont {S.~A.}\ \bibnamefont {Bender}}, \bibinfo {author} {\bibfnamefont {L.}~\bibnamefont {Baldrati}}, \bibinfo {author} {\bibfnamefont {F.}~\bibnamefont {Kronast}}, \bibinfo {author} {\bibfnamefont {A.}~\bibnamefont {Qaiumzadeh}}, \bibinfo {author} {\bibfnamefont {J.}~\bibnamefont {Sinova}}, \bibinfo {author} {\bibfnamefont {A.}~\bibnamefont {Brataas}}, \bibinfo {author} {\bibfnamefont {R.~A.}\ \bibnamefont {Duine}},\ and\ \bibinfo {author} {\bibfnamefont {M.}~\bibnamefont {Kläui}},\ }\bibfield  {title} {\bibinfo {title} {Anisotropies and magnetic phase transitions in insulating antiferromagnets determined by a {Spin}-{Hall} magnetoresistance probe},\ }\href {https://doi.org/10.1038/s42005-019-0150-8} {\bibfield  {journal} {\bibinfo  {journal} {Commun Phys}\ }\textbf {\bibinfo
  {volume} {2}},\ \bibinfo {pages} {1} (\bibinfo {year} {2019})}\BibitemShut {NoStop}%
\bibitem [{\citenamefont {Wang}\ \emph {et~al.}(2021)\citenamefont {Wang}, \citenamefont {Xiao}, \citenamefont {Guo}, \citenamefont {Lee-Wong}, \citenamefont {Yan}, \citenamefont {Cheng},\ and\ \citenamefont {Du}}]{wang_spin_2021}%
  \BibitemOpen
  \bibfield  {author} {\bibinfo {author} {\bibfnamefont {H.}~\bibnamefont {Wang}}, \bibinfo {author} {\bibfnamefont {Y.}~\bibnamefont {Xiao}}, \bibinfo {author} {\bibfnamefont {M.}~\bibnamefont {Guo}}, \bibinfo {author} {\bibfnamefont {E.}~\bibnamefont {Lee-Wong}}, \bibinfo {author} {\bibfnamefont {G.~Q.}\ \bibnamefont {Yan}}, \bibinfo {author} {\bibfnamefont {R.}~\bibnamefont {Cheng}},\ and\ \bibinfo {author} {\bibfnamefont {C.~R.}\ \bibnamefont {Du}},\ }\bibfield  {title} {\bibinfo {title} {Spin {Pumping} of an {Easy}-{Plane} {Antiferromagnet} {Enhanced} by {Dzyaloshinskii}–{Moriya} {Interaction}},\ }\href {https://doi.org/10.1103/PhysRevLett.127.117202} {\bibfield  {journal} {\bibinfo  {journal} {Phys. Rev. Lett.}\ }\textbf {\bibinfo {volume} {127}},\ \bibinfo {pages} {117202} (\bibinfo {year} {2021})}\BibitemShut {NoStop}%
\bibitem [{\citenamefont {Vélez}\ \emph {et~al.}(2019)\citenamefont {Vélez}, \citenamefont {Schaab}, \citenamefont {Wörnle}, \citenamefont {Müller}, \citenamefont {Gradauskaite}, \citenamefont {Welter}, \citenamefont {Gutgsell}, \citenamefont {Nistor}, \citenamefont {Degen}, \citenamefont {Trassin}, \citenamefont {Fiebig},\ and\ \citenamefont {Gambardella}}]{velez_high-speed_2019}%
  \BibitemOpen
  \bibfield  {author} {\bibinfo {author} {\bibfnamefont {S.}~\bibnamefont {Vélez}}, \bibinfo {author} {\bibfnamefont {J.}~\bibnamefont {Schaab}}, \bibinfo {author} {\bibfnamefont {M.~S.}\ \bibnamefont {Wörnle}}, \bibinfo {author} {\bibfnamefont {M.}~\bibnamefont {Müller}}, \bibinfo {author} {\bibfnamefont {E.}~\bibnamefont {Gradauskaite}}, \bibinfo {author} {\bibfnamefont {P.}~\bibnamefont {Welter}}, \bibinfo {author} {\bibfnamefont {C.}~\bibnamefont {Gutgsell}}, \bibinfo {author} {\bibfnamefont {C.}~\bibnamefont {Nistor}}, \bibinfo {author} {\bibfnamefont {C.~L.}\ \bibnamefont {Degen}}, \bibinfo {author} {\bibfnamefont {M.}~\bibnamefont {Trassin}}, \bibinfo {author} {\bibfnamefont {M.}~\bibnamefont {Fiebig}},\ and\ \bibinfo {author} {\bibfnamefont {P.}~\bibnamefont {Gambardella}},\ }\bibfield  {title} {\bibinfo {title} {High-speed domain wall racetracks in a magnetic insulator},\ }\href {https://doi.org/10.1038/s41467-019-12676-7} {\bibfield  {journal} {\bibinfo  {journal} {Nat Commun}\ }\textbf {\bibinfo
  {volume} {10}},\ \bibinfo {pages} {4750} (\bibinfo {year} {2019})}\BibitemShut {NoStop}%
\bibitem [{\citenamefont {Cogulu}\ \emph {et~al.}(2021)\citenamefont {Cogulu}, \citenamefont {Statuto}, \citenamefont {Cheng}, \citenamefont {Yang}, \citenamefont {Chopdekar}, \citenamefont {Ohldag},\ and\ \citenamefont {Kent}}]{cogulu_direct_2021}%
  \BibitemOpen
  \bibfield  {author} {\bibinfo {author} {\bibfnamefont {E.}~\bibnamefont {Cogulu}}, \bibinfo {author} {\bibfnamefont {N.~N.}\ \bibnamefont {Statuto}}, \bibinfo {author} {\bibfnamefont {Y.}~\bibnamefont {Cheng}}, \bibinfo {author} {\bibfnamefont {F.}~\bibnamefont {Yang}}, \bibinfo {author} {\bibfnamefont {R.~V.}\ \bibnamefont {Chopdekar}}, \bibinfo {author} {\bibfnamefont {H.}~\bibnamefont {Ohldag}},\ and\ \bibinfo {author} {\bibfnamefont {A.~D.}\ \bibnamefont {Kent}},\ }\bibfield  {title} {\bibinfo {title} {Direct imaging of electrical switching of antiferromagnetic {N\'eel} order in {$\alpha$-Fe$_2$O$_3$} epitaxial films},\ }\href {https://doi.org/10.1103/PhysRevB.103.L100405} {\bibfield  {journal} {\bibinfo  {journal} {Phys. Rev. B}\ }\textbf {\bibinfo {volume} {103}},\ \bibinfo {pages} {L100405} (\bibinfo {year} {2021})}\BibitemShut {NoStop}%
\bibitem [{\citenamefont {Zhang}\ and\ \citenamefont {Cheng}(2022)}]{zhang2022theory}%
  \BibitemOpen
  \bibfield  {author} {\bibinfo {author} {\bibfnamefont {H.}~\bibnamefont {Zhang}}\ and\ \bibinfo {author} {\bibfnamefont {R.}~\bibnamefont {Cheng}},\ }\bibfield  {title} {\bibinfo {title} {Theory of harmonic hall responses of spin-torque driven antiferromagnets},\ }\href@noop {} {\bibfield  {journal} {\bibinfo  {journal} {Journal of Magnetism and Magnetic Materials}\ }\textbf {\bibinfo {volume} {556}},\ \bibinfo {pages} {169362} (\bibinfo {year} {2022})}\BibitemShut {NoStop}%
\bibitem [{\citenamefont {Qaiumzadeh}\ \emph {et~al.}(2017)\citenamefont {Qaiumzadeh}, \citenamefont {Skarsv\aa{}g}, \citenamefont {Holmqvist},\ and\ \citenamefont {Brataas}}]{SpinSuperfluidAFM}%
  \BibitemOpen
  \bibfield  {author} {\bibinfo {author} {\bibfnamefont {A.}~\bibnamefont {Qaiumzadeh}}, \bibinfo {author} {\bibfnamefont {H.}~\bibnamefont {Skarsv\aa{}g}}, \bibinfo {author} {\bibfnamefont {C.}~\bibnamefont {Holmqvist}},\ and\ \bibinfo {author} {\bibfnamefont {A.}~\bibnamefont {Brataas}},\ }\bibfield  {title} {\bibinfo {title} {Spin superfluidity in biaxial antiferromagnetic insulators},\ }\href {https://doi.org/10.1103/PhysRevLett.118.137201} {\bibfield  {journal} {\bibinfo  {journal} {Phys. Rev. Lett.}\ }\textbf {\bibinfo {volume} {118}},\ \bibinfo {pages} {137201} (\bibinfo {year} {2017})}\BibitemShut {NoStop}%
\end{thebibliography}%

\end{document}